\documentclass[conference]{IEEEtran}
\IEEEoverridecommandlockouts
\usepackage{cite}
\usepackage{braket}
\usepackage{amsmath,amssymb,amsfonts} 
\usepackage{algorithm}
\usepackage[utf8]{inputenc}
\usepackage{array}
\usepackage{colortbl}
\usepackage{tabularray}
\usepackage{rotating}
\usepackage{color}
\usepackage{svg}

\usepackage{graphicx}
\usepackage{textcomp}
\usepackage{xcolor}
\usepackage{gensymb}
\usepackage{subfigure}
\usepackage{epsfig}
\usepackage{epstopdf}
\usepackage{array}
\usepackage{makecell} 
\usepackage{makecell}
\usepackage{textcomp}
\usepackage{booktabs}
\usepackage{threeparttable}
\usepackage{booktabs}
\usepackage{caption}
\usepackage{multirow}
\usepackage{algorithm}
\usepackage{algpseudocode}
\usepackage{amsmath}

\def\BibTeX{{\rm B\kern-.05em{\sc i\kern-.025em b}\kern-.08em
    T\kern-.1667em\lower.7ex\hbox{E}\kern-.125emX}}
\begin{document}

\title{Semantic-driven Wireless Environment Knowledge Representation for Efficiency-Accuracy Balanced Beam Prediction in Vehicular Networks
\\
}

\author{\IEEEauthorblockN{Jialin Wang$^*$, Jianhua Zhang$^\dag $, Yu Li$^\dag $, Yutong Sun$^*$, Yuxiang Zhang$^\dag $}
\IEEEauthorblockA{
$^*$ China Information Technology Designing \& Consulting Institute Co., Ltd., Beijing, 100048, China \\
$^\dag $ State Key Lab of Networking and Switching Technology,
Beijing University of Posts and Telecommunications, \\
Beijing, 100876, China,\\
$^*$ Email:\{wangjl733, sunyt79\}@chinaunicom.cn
$^\dag $ Email:\{jhzhang, li.yu, zhangyx\}@bupt.edu.cn\\
}
}

\maketitle

\begin{abstract}

The rapid evolution of the internet of vehicles demands ultra-reliable low-latency communication in high-mobility environments, where conventional beam prediction methods suffer from high-dimensional inputs, prolonged training times, and limited interpretability. To address these challenges, the propagation environment semantics-aware wireless environment knowledge beam prediction (PES-WEKBP) framework is proposed.
PES-WEKBP pioneers a novel electromagnetic (EM)-grounded knowledge distillation method, transforming raw visual data into an ultra-lean, interpretable material and location-related wireless environment knowledge matrix. 
This matrix explicitly encodes critical propagation environment semantics, which is material EM properties and spatial relationships through a physics-informed parameterization process, distilling the environment and channel interplay into a minimal yet information-dense representation.
A lightweight decision network then leverages this highly compressed knowledge for low-complexity beam prediction.
To holistically evaluate the performance of PES-WEKBP, we first design the prediction consistency-efficiency index (PCEI), which combines prediction accuracy with a stability-penalized logarithmic training time to ensure a balanced optimization of reliability and computational efficiency. 
Experiments validate that PES-WEKBP achieves a 99.75\% to 99.96\% dimension reduction and improves accuracy by 5.52\% to 8.19\%, which outperforms state-of-the-art methods in PCEI scores across diverse vehicular scenarios. 

\end{abstract}


\section{Introduction}
Driven by the convergence of emerging internet of things (IoT) techniques \cite{lakhan2024augmented,zhang2024wireless}, advanced sensing capabilities \cite{almehdhar2024deep,11103469,8889718,wang2024digital}, intelligent systems \cite{zhang2022toward,10472706}, and novel wireless communication techniques like integrated sensing and communication (ISAC) \cite{yu2025channelgpt,zhang2023integrated,9921271}, future internet of vehicles (IoV) systems are expected to acquire multi-modal environmental awareness and achieve intelligent coordination across communication links \cite{peng2018vehicular,wang2024clutter,zhang20183d}.
The realization of this vision hinges on accurate and efficient beam prediction to establish reliable communication links.
Yet, a critical challenge emerges: the inherent trade-off between prediction accuracy and computational efficiency.  
Next-generation vehicular applications such as autonomous platooning require ultra-low latency below 10 ms with 99.999\% reliability, while high-definition map updates demand centimeter-level precision within strict time constraints. 
Conventional data-driven beam prediction methods, especially those based on high-dimensional raw sensing data, often fail to simultaneously meet both demands in high-mobility and complex scenarios.
Consequently, balancing accuracy and efficiency has become a fundamental issue for enabling real-time beam management in intelligent vehicular networks.

Conventional beam selection methods typically rely on pilot training and channel state information (CSI) feedback, where the optimal beam is selected from predefined beam codebooks \cite{dreifuerst2023machine,qi2021acquisition,zheng2022survey}.
These strategies, however, introduce substantial training overhead and lack adaptability in dynamic non-line-of-sight (NLoS) environments where channels change rapidly \cite{shi2025can,zhang20173}. 
Recent advances incorporating sensing technologies and artificial intelligence (AI) have led to environment-based beam prediction, which leverages deep neural networks (DNNs) to learn environment-channel mappings from multi-modal data. Examples include image-based and camera-assisted beam prediction \cite{zhong2024image,charan2024camera}, vision-aided selection using image segmentation \cite{wen2023vision}, and beam selection utilizing propagation environment semantics (PES) \cite{sun2023define,sun2023pc}.
These methods attempt to improve prediction accuracy in complex IoV scenarios by fusing multi-source sensing data, but several challenges remain unresolved.

On one hand, naive data-level fusion such as direct concatenation of high-dimensional raw data suffers from the curse of dimensionality\cite{hu2024tackling} and obscures radio propagation-critical features within irrelevant information.
On the other hand, although some dimension-reduction techniques, such as image segmentation or semantic extraction, improve efficiency, their feature construction often follows computer vision logic rather than incorporating explicit electromagnetic (EM) propagation principles.
This forces DNNs to implicitly learn physical laws, resulting in excessive model complexity and inference latency.
Collectively, these issues hinder the balance between prediction accuracy and computational efficiency in complex IoV scenarios. 

To address these challenges, this paper advocates a paradigm shift: from purely data-driven learning to constructing explainable environment-beam associations grounded in EM theory.
Inspired by the recently proposed wireless environment knowledge pool (WEKP) framework \cite{10829758}, which constructs interpretable relationships between environment and channel through analytical models and relational reasoning, we develop a novel semantic-driven representation for wireless environment knowledge (WEK). 
Specifically, WEKP serves as a self-evolving repository capable of storing and refining environmental insights. 
When sensing data is input, it retrieves relevant knowledge regarding channel characteristics and communication contexts, thereby simplifying or even bypassing explicit CSI acquisition.
The significant potential of WEKP-based channel prediction and channel digital twinning has been demonstrated in existing studies \cite{wang2024electromagnetic,cui2025overview,basaran2025gen}.

In this paper, we propose a Propagation Environment Semantics-based Wireless Environment Knowledge representation and Beam Prediction (PES-WEKBP) framework. 
The core idea is to use semantically abstracted propagation environment representations to drive the construction of physically meaningful features for beam prediction. 
Whereas PES offers an efficient and compact description of the propagation scene, WEK incorporates deeper EM-domain knowledge, such as wave-material interaction mechanisms, forming a structured knowledge representation that supports efficient and accurate beam prediction.
The main contributions are summarized as follows:

\begin{itemize}
    \item We propose a novel framework for interpretable environment-beam mapping based on EM principles. Its core is an image-based WEK representation principle, extending WEK to 2D imagery and enabling dimensionality reduction via multi-scale image segmentation.
    
    \item Building on this foundation, we propose a segmentation-based PES extraction method that transforms complex images into concise semantic descriptors.  It merges or eliminates propagation-irrelevant features, reducing environmental category redundancy by at least 60\% across diverse traffic scenarios.
    
    \item Leveraging the extracted PES, we design a material and location-aware WEK representation that explicitly integrates material-specific propagation traits and distance-based relationships. This method achieves a  99.75\% to 99.96\% input dimensionality reduction and 82.70\% reduced inference latency.

    \item We propose a performance-cost efficiency index (PCEI) that combines prediction accuracy with logarithmic time complexity and an exponential decay factor. Results show that PES-WEKBP confirms ts balanced optimization of accuracy gains ranging from 5.52\% to 8.19\% alongside computational cost reduction exceeding 99\%.

\end{itemize}


\section{System Model and Problem Formulation}\label{secII}

\begin{figure}[!t]
	\centering
	\includegraphics[scale=0.35]{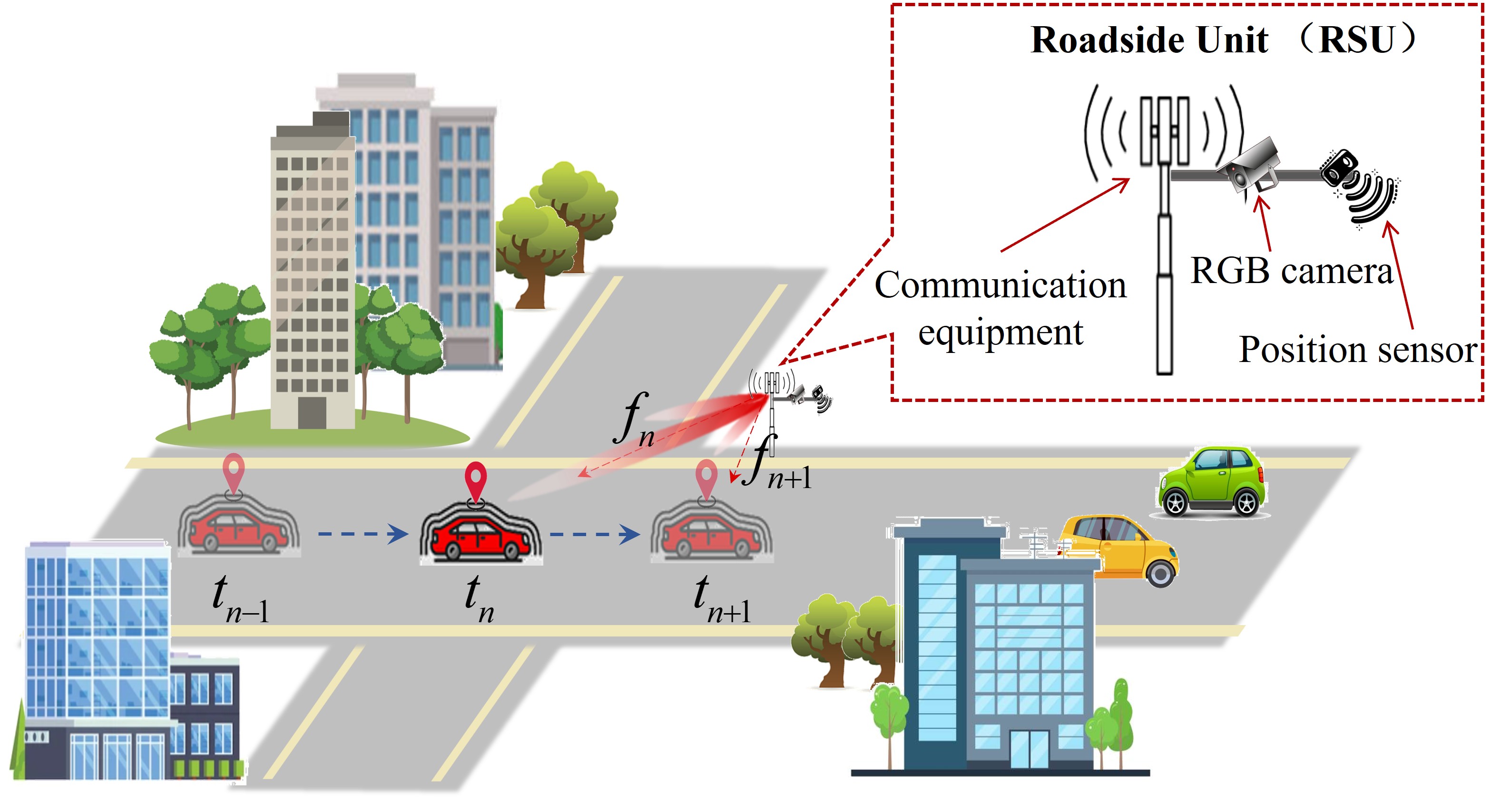}
	\caption{Scenario diagram of vehicle-connected V2I communication based on environment awareness.}\label{SceneGraph}
\end{figure}

\subsection{System and Channel Model}

\begin{figure*}[!t]
	\centering
	\includegraphics[scale=0.37]{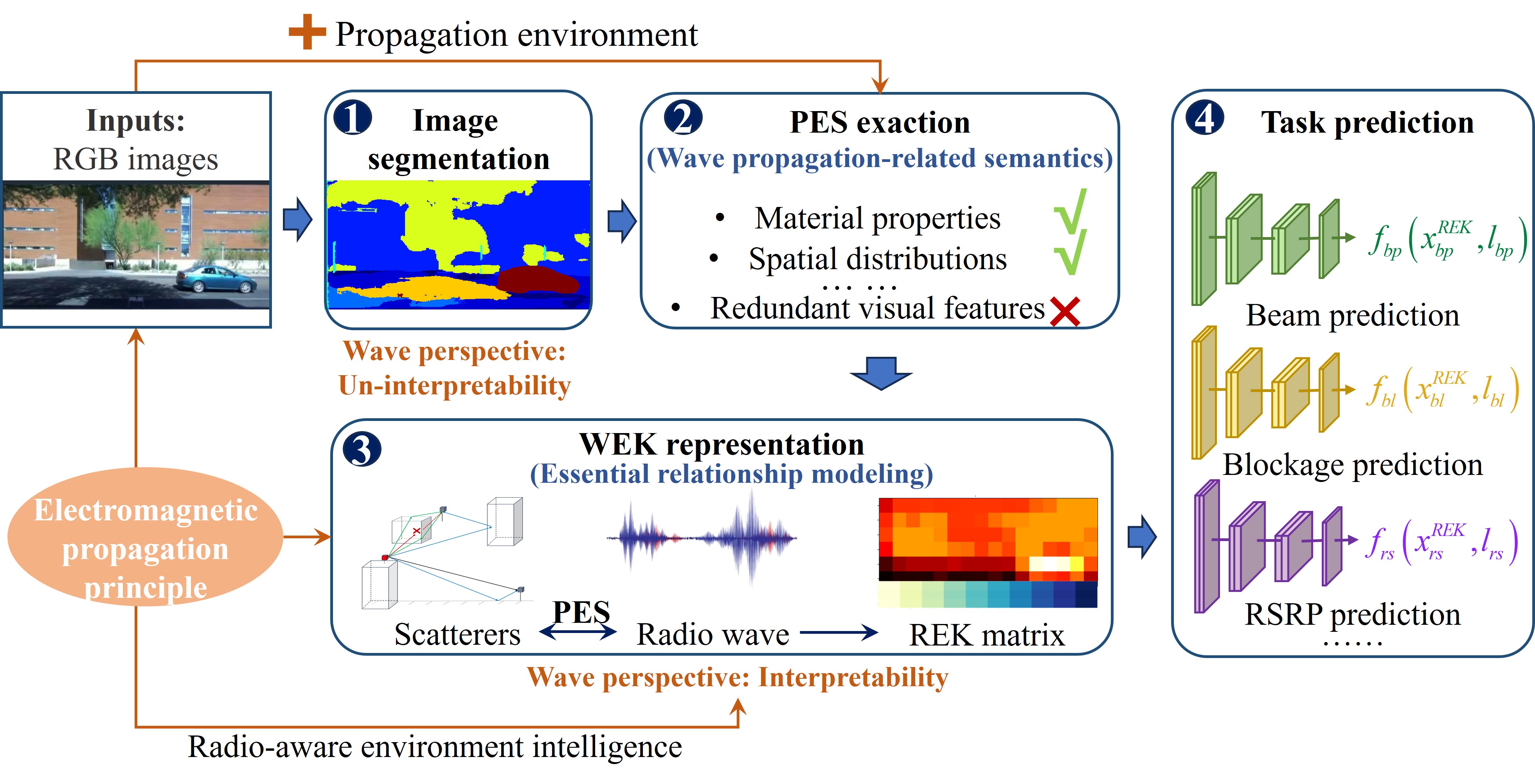}
	\caption{PES-WEKBP framework, consisting image segmentation, PES exaction, REK representation and task prediction,}\label{SystDiag}
\end{figure*}

Consider a downlink urban road scenario with continuous coverage by a roadside unit (RSU) over two unidirectional lanes. The RSU and vehicles communicate via millimeter-wave (mmWave) vehicles to infrastructure (V2I) links, employing beamforming to compensate for mmWave path loss. As shown in Fig. \ref{SceneGraph}, the RSU is equipped with a mmWave massive multiple-input multiple-output (MIMO) antenna array, an RGB camera, and LiDAR sensors. The transmitter (Tx) uses a phased array operating in the mmWave band with $N_t$ antennas. The base station (BS) receives signals via predefined oversampled beam codebooks. Mobile users (vehicles) are equipped with single-antenna mobile vehicle units (MVUs) for omnidirectional mmWave transmission. The mmWave channel between the BS and mobile user is modeled as:
\begin{equation}
h[k] = \sum_{l} a_l e^{-j2\pi f_D k \pi t + j \phi_l} a(\theta_{l,az}, \theta_{l,el}),
\end{equation}
where $L$ is the number of multipath components, $f_D$ is the Doppler frequency, $a_l$, $\phi_l$, and $\tau_l$ are the gain, phase shift, and delay of the $l$-th path, and $a(\theta_{l,az}, \theta_{l,el})$ is the array response vector with azimuth ($\theta_{l,az}$) and elevation ($\theta_{l,el}$) angles.

The mmWave BS employs analog beamforming with phase shifters. The beamforming vector $\mathbf{w}$ is selected from a codebook $\mathcal{F} = \{f_1, \ldots, f_B\}$, where the $n$-th beamforming vector is defined as:
\begin{equation}
{f_n} = \frac{1}{{{N_t}}}{\left[ {1, \ldots ,{e^{j\beta ({N_t} - 1)\sin ({\theta _{i,el}})\cos ({\theta _{i,az}})}}} \right]^T},
\end{equation}
where $\beta = \frac{2\pi d f_D k}{c}$, $d$ is antenna spacing, and $c$ is the speed of light. The optimal beamforming vector $f^{\text{opt}}$ maximizes the achievable rate:
\begin{equation}
\text{Rate}_{f} = \frac{1}{K} \sum_{k=1}^{K} \log_{2} \left( 1 + \frac{P_{k}}{\sigma^{2}} |h^{T}[k] \mathcal{F}|^2 \right),
\end{equation}
\begin{equation}
f^{\text{opt}} = \arg \max_{f \in \mathcal{F}} \text{Rate}_{f}.
\end{equation}

In mmWave communication V2I scenarios, the system model, channel model, and implementation process of beam prediction tasks are tightly coupled. The system model involves a base station equipped with a massive MIMO system serving multiple single-antenna users, where the received signal $ y_k $ is determined by the channel $ \mathbf{h}_k $, beamforming vector $ \mathbf{f} $, transmitted signal $ s_k $, and additive white Gaussian noise $ \epsilon_k $. The channel $ \mathbf{h}_k $ is characterized by multipath propagation, with each path having distinct gain, phase shift, delay, and angle of arrival. To maximize transmission rates, the system selects the optimal beamforming vector $ \mathbf{f}_{\text{opt}} $ from a predefined codebook $ \mathcal{F} $, traditionally relying on pilot-based channel estimation. However, in high-mobility environments, such methods face impractical latency issues. To overcome this, a semantic environment-aware approach is employed to predict future channel fading states by integrating environmental image semantics and user-infrastructure positional data. This approach constructs a wireless environment knowledge representation that explicitly models implicit relationships between EM wave propagation and environmental features, reduces information redundancy, and utilizes simplified neural architectures to achieve accurate channel prediction. Consequently, beamforming strategies are optimized, significantly enhancing mmWave communication performance.

\subsection{Image-Based WEK Representation}

This paper proposes an image-based WEK representation principle, establishing a mapping relationship that converts visual information within the Tx and receiver (Rx) coverage area into feature vectors correlated with channel knowledge. By analyzing environmental attributes in images, this methodology extracts implicit radio propagation information and constructs interpretable, quantifiable environment-channel relationships.

In computer vision-assisted communication research, RGB images are predominantly adopted due to their universality and acquisition feasibility, enabling convenient deployment of RGB sensors at BS or RSUs. An RGB image constitutes a three-dimensional tensor $ P \in \mathbb{R}^{H \times W \times 3} $, where $ H $ and $ W $ denote height and width, respectively. Each pixel's intensity is expressed as:
\begin{equation}
P(h,w) = [R(h,w), G(h,w), B(h,w)],
\end{equation}
where $ R(h,w) $, $ G(h,w) $, and $ B(h,w) $ represent the intensity values of the red, green, and blue channels at position $ (h,w) $.

Radio propagation is influenced by environmental factors including building distributions, user mobility patterns, and obstructions like vegetation \cite{zhang2024wireless,sun2022environment}. 
Initial processing utilizes image analysis techniques such as edge detection and deep learning to preserve dominant objects along the transmission-reception pathway. This stage extracts essential characteristics including contours, volumetric properties, positional coordinates, and height ratios between transmitters and receivers.
Irrelevant objects outside the communication link are eliminated. A feature extraction function $ E $ maps the RGB image to an environmental feature space $ F = E(P) $, where $ E $ is typically implemented via deep learning for preliminary propagation-relevant feature screening.

\begin{figure*}[!t]
	\centering
	\includegraphics[scale=0.3]{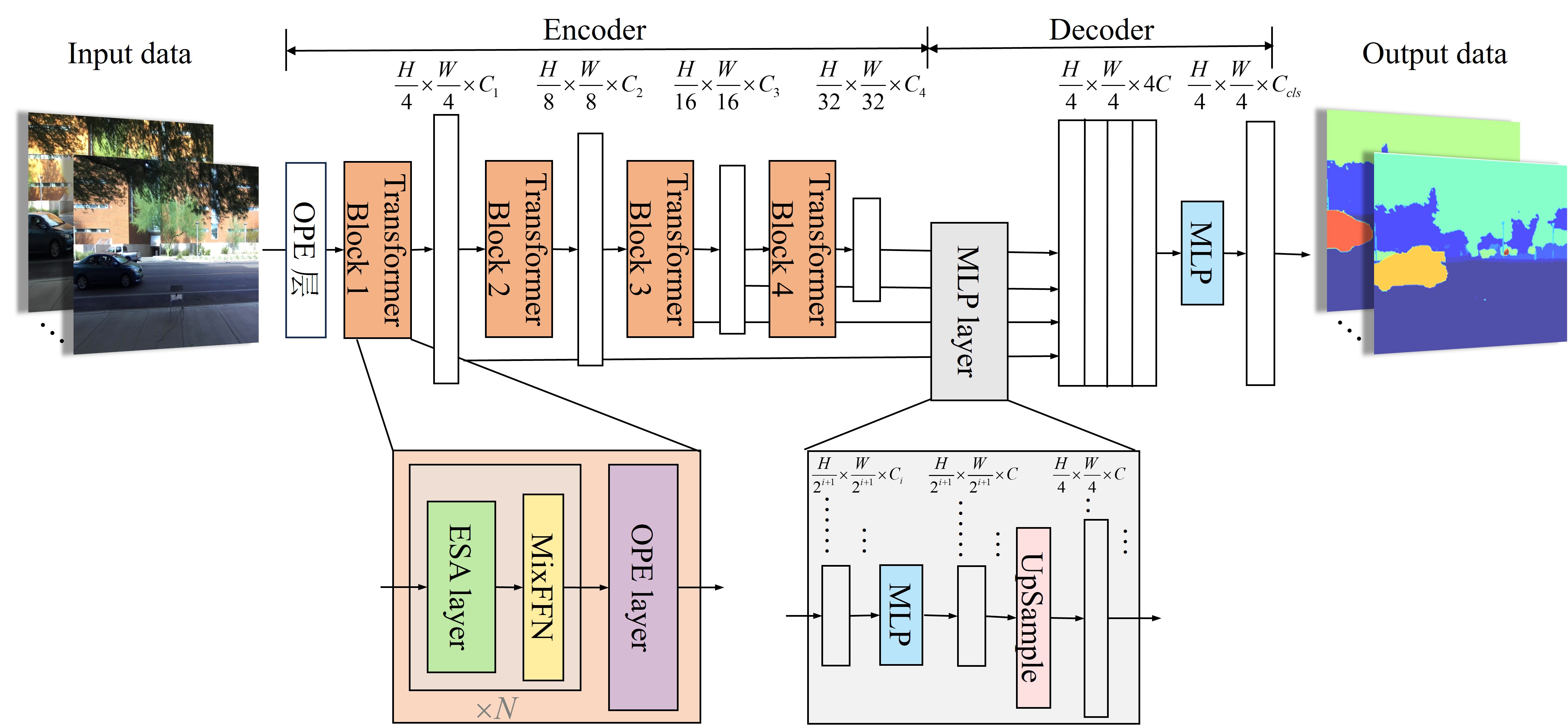}
	\caption{Seg-TIS network architecture.}\label{SegformerNet}
\end{figure*}

Let $ F(t) $ denote the environmental features extracted from images captured by the RSU at time $ t $, and $ P(t) $ represent the updated image information. The wireless channel and its associated knowledge $ Z(t) $ are fully determined by radio wave properties, the updated image $ P(t) $, and propagation environment features $ F(t) $. Thus, the image-based WEK representation process for a specific radio wave can be formulated as:
\begin{equation}
Z(t) = \mathcal{K}(F(t), P(t)),
\end{equation}
where $ \mathcal{K}(\cdot, \cdot) $ is a function that maps the updated image and propagation features to channel knowledge. 
However, directly constructing $ \mathcal{K}(\cdot, \cdot) $ faces two key challenges:
On the one hand, the interaction between radio waves and the propagation environment is highly nonlinear and difficult to model with universally applicable functions.
On the other hand, propagation features exhibit strong scenario-dependence, particularly due to factors such as terrain variations, structural geometries, and dielectric properties, which limits cross-scenario generalizability.

To address this, we extract radio-aware image features $ W(t) $ encoding propagation information beyond visual semantics. Under short-term static conditions where $ P(t) \approx P $ for $ 0 \leq t \leq T $, channel knowledge simplifies to:
\begin{equation}
Z(t) = \mathcal{K}(W(t), P),
\end{equation}
where $ W(t) $ is derived through EM theory-guided image segmentation. When quantifiable environment-knowledge relationships exist, image reconstruction becomes feasible via:
\begin{equation}
P(t) = \mathcal{G}(W, Z).
\end{equation}

This enables dynamic WEK updates through historical feature-channel knowledge synthesis. Our core objective is to establish theoretically grounded, quantifiable mappings $ \mathcal{K}(\cdot,\cdot) $ between environmental characteristics and channel states.

\section{Semantic-driven WEK Representation for Beam Prediction}\label{secIII}

This section details the proposed PES-WEKBP framework, consisting image segmentation, PES exaction, knowledge representation and task prediction, as shown in Fig. \ref{SystDiag}. 
Unlike conventional methods that rely on direct application of segmented image semantics such as vehicle and road classifications for communication tasks, PES-WEKBP introduces a novel semantics paradigm.  
The core innovation lies in its physics-driven knowledge representation, which deconstructs the environment from an EM propagation perspective and explicitly models the intrinsic relationships between scatterers and radio waves.  Specifically, PES extracts semantics critical to wave propagation, such as material properties and spatial distributions, while filtering redundant visual features that are not related to radio propagation behavior.  
This approach bridges the gap between human-interpretable image semantics and radio-aware environment intelligence, achieving significant dimensionality reduction without sacrificing task accuracy.


\subsection{Segformer-Based Traffic Image Segmentation Method}
In V2I communication scenarios, RGB cameras at RSUs provide high-resolution traffic images with broad coverage and continuous data acquisition, yet face challenges including occlusion-induced information loss, motion blur from rapid vehicle movement, cluttered backgrounds, perspective distortion, and limited field-of-view \cite{ojala2019novel,zhang2024considering}.  
These defects degrade segmentation accuracy and hinder robust environmental sensing.

The Segformer network addresses such issues through hierarchical Transformer encoders that generate multi-scale features and an multilayer perceptron (MLP) decoder for cross-layer fusion, leveraging local-global attention to achieve superior semantic segmentation\cite{xie2021segformer}.  
For instance, in IoV-oriented semantic communication tasks, a Segformer-based framework achieved 80.7\% mean accuracy and 96.6\% Rank-1 accuracy in traffic vehicle recognition, outperforming conventional methods\cite{yu2023semantic}.  Inspired by this, we propose the segformer-based traffic image segmentation (Seg-TIS) method, which adapts segformer’s architecture to V2I scenarios by enhancing occlusion robustness, motion blur tolerance, and clutter suppression.  

The Segformer architecture employs a hierarchical Transformer encoder and a lightweight MLP decoder, as illustrated in Fig. \ref{SegformerNet}. The encoder first divides the input image into $4 \times 4$ patches and processes them through overlapping patch embedding. A series of Mix Transformer (MiT) blocks then extract multi-scale features at resolutions from 1/4 to 1/32 of the original image. Each block comprises an efficient self-attention (ESA) mechanism for capturing global context and a mixed feedforward network (MixFFN) module for local feature transformation.
The MixFFN layer omits explicit positional encoding and instead integrates a $3 \times 3$ convolutional layer between two linear transformation layers to fuse spatial information:
\begin{equation}
x_{\text{out}} = \text{MLP}\left(\text{ReLU}\left(\text{Conv}_{3\times3}(x_{\text{in}})\right)\right) + x_{\text{in}},
\end{equation}
where $ x_{\text{in}} $ is the input feature from the self-attention module, $ \text{Conv}_{3\times3}(\cdot) $ denotes a $3 \times 3$ convolution, $ \text{ReLU}(\cdot) $ is the activation function, and $ x_{\text{out}} $ is the output feature.

The encoder outputs multi-scale feature maps $ F_i \in \mathbb{R}^{C_i \times \frac{H}{2} \times \frac{W}{2}} $ ($ i = 1, 2, 3, 4 $), where $ F_i $ represents the feature map at the $ i $-th scale, $ C_i $ is the number of channels, and $ \frac{H}{2} \times \frac{W}{2} $ denotes the spatial resolution. These feature maps encapsulate hierarchical information ranging from coarse to fine details, providing rich context for semantic segmentation.

The Seg-TIS decoder, entirely composed of MLP layers, avoids computationally intensive components typical of traditional decoders. Its simplicity leverages the expansive effective receptive fields provided by the hierarchical Transformer encoder. The decoding process first unifies the channel dimensions of multi-level features through MLP layers, upsamples them to 1/4 of the original resolution, concatenates them, and then uses another MLP layer to fuse the concatenated features. Finally, a final MLP layer converts the fused features into the predicted segmentation mask $ M $, with dimensions $ H/4 \times W/4 \times N_{\text{cls}} $, where $ N_{\text{cls}} $ is the number of segmentation categories. The $ (i, j) $-th entry of $ M $ represents the category label at pixel coordinate $ (i, j) $, i.e., $ M(i, j) = 0, \ldots, N_{\text{cls}} - 1 $. The loss function for Seg-TIS is defined as:
\begin{equation}
\begin{array}{l}
{L_{{\rm{(Seg - TIS)}}}} =  - \frac{1}{{H \times W}} \times \sum\limits_{n = 1}^{{N_{{\rm{cls}}}}} {\sum\limits_{i = 1}^{H/4} {} } \\
\sum\limits_{j = 1}^{W/4} {\left[ {{Y_{h,w}}\log \left( {{M_{h,w}}} \right) + \left( {1 - {Y_{h,w}}} \right)\log \left( {1 - {M_{h,w}}} \right)} \right]} 
\end{array},
\end{equation}
where $ M_{h,i,j} $ is the predicted probability of pixel $ (i, j) $ belonging to category $ n $, and $ Y_{h,i,j} $ is the ground truth probability. The network is optimized using stochastic gradient descent until convergence. For each pixel $ (i, j) $, the category with the highest probability is selected as the segmentation result:
\begin{equation}
S_{h,w} = \arg \max_{i=0,\ldots,N_{\text{cls}}-1} M_{i,h,w}.
\end{equation}

After obtaining the segmentation map $ S_{h,w} $, binary mask images are generated for distinct semantic classes including vegetation, buildings, and terrain by assigning pixel values of interest to 255 while setting all others to 0.

\subsection{Semantic Extraction Method Based on RGB Images}
In the first module of the PES-WEKBP architecture, a series of binary mask images are obtained, which are primarily interpreted from a human visual perspective. However, in wireless propagation processes, the degree of impact of different categories on EM waves varies significantly. Some categories exhibit consistent effects on EM waves during propagation, while others exert substantial influence, and some categories have almost no effect.

\begin{figure}[!t]
	\centering
	\includegraphics[scale=0.45]{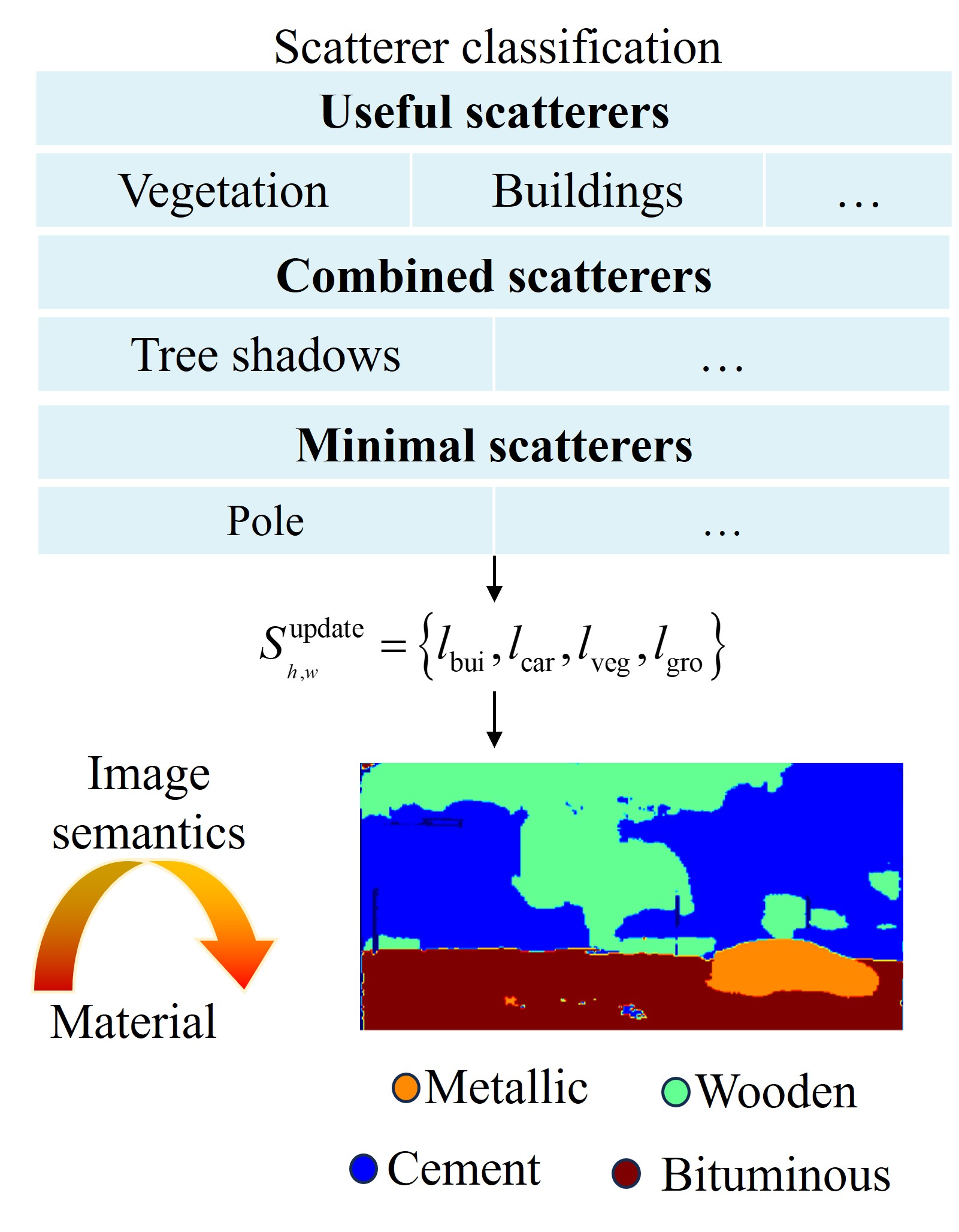}
	\caption{The proposed PES extraction method.}\label{PESextraction}
\end{figure}

To bridge the gap between computer vision/human cognitive perspectives and radio propagation perspectives for image interpretation, this paper proposes a PES extraction method based on segmented images, as shown in Fig. \ref{PESextraction}. The segmentation map $ S_{h,w} $ output from the Seg-TIS network can be represented as a set containing all categories: $ S_{h,w} = C_{\text{use}} \cup C_{\text{com}} \cup C_{\text{min}} $. 
Here, $ C_{\text{use}} = \{ l_{u1}, l_{u2}, \ldots, l_{vw} \} $ denotes the set of category labels that significantly impact EM wave propagation, including objects or regions critical to propagation characteristics, such as buildings, vegetation, and moving vehicles. $ C_{\text{com}} = \{ l_{c1}, l_{c2}, \ldots, l_{vw} \} $ represents a subset of segmentation map categories relevant to EM wave propagation. 
For example, from a human visual perspective, the ground can be subdivided into elements like sidewalks, roads, and shadows formed by sunlight passing through vegetation and buildings, while moving vehicles are categorized into cars, trucks, buses, motorcycles, etc.

These elements and vehicle types are effectively identified as distinct binary masks during image segmentation. However, from an EM wave propagation perspective, this information corresponds to ground reflections and vehicles in the propagation process. 
Ground surfaces with different social attributes and their coverings have negligible effects on EM wave propagation and can thus be ignored. Similarly, different vehicle types essentially produce metal reflections, allowing vehicle-type distinctions to be disregarded. 
Categories with minimal or no significant impact on propagation are collectively denoted as $ C_{\text{min}} = \{ l_{c1}, l_{c2}, \ldots, l_{vw} \} $. 
By analyzing propagation environment information from an EM wave perspective, the final segmentation map categories merge similar classes into new ones and retain critical categories while removing insignificant ones, i.e., $ C_{\text{update}} = C_{\text{use}} \cup C_{\text{com}} \setminus C_{\text{min}} $. 
In this work, sidewalks, roads, and object shadows are fused and reconstructed into a unified "ground" pixel classification label $ l_{\text{go}} $, while multiple vehicle types are merged into a unified “vehicle” pixel classification label $ l_{\text{veh}} $. Additionally, segmentation map categories with significant impact, such as vegetation $ l_{\text{veg}} $ and buildings $ l_{\text{bui}} $, are retained. All other categories are deemed insignificant. Based on this, the segmentation map labels $ S^{\text{update}}_{h,w} = \{ l_{\text{bui}}, l_{\text{veh}}, l_{\text{veg}}, l_{\text{go}} \} $ are converted into corresponding material labels, completing the propagation environment semantic extraction based on segmented images. This is expressed as:
\begin{equation}
T^{\text{material}}_{h,w} = M\left( S^{\text{update}}_{h,w} \right) = \left\{ M(l_{\text{bui}}), M(l_{\text{veh}}), M(l_{\text{veg}}), M(l_{\text{go}}) \right\}.
\end{equation}
where $ M(\cdot) $ maps segmentation categories to material types. In this work, four material classifications are considered: cement for buildings, metal for vehicles, wood for vegetation, and asphalt for ground. After propagation environment semantic extraction, the images undergo secondary segmentation based on EM wave propagation principles. These binary mask images are differentiated by pixel values annotated with material categories, though the values themselves lack inherent weight significance. 

\subsection{Material and Location-Related Knowledge Representation}

\begin{figure}[!t]
	\centering
	\includegraphics[scale=0.45]{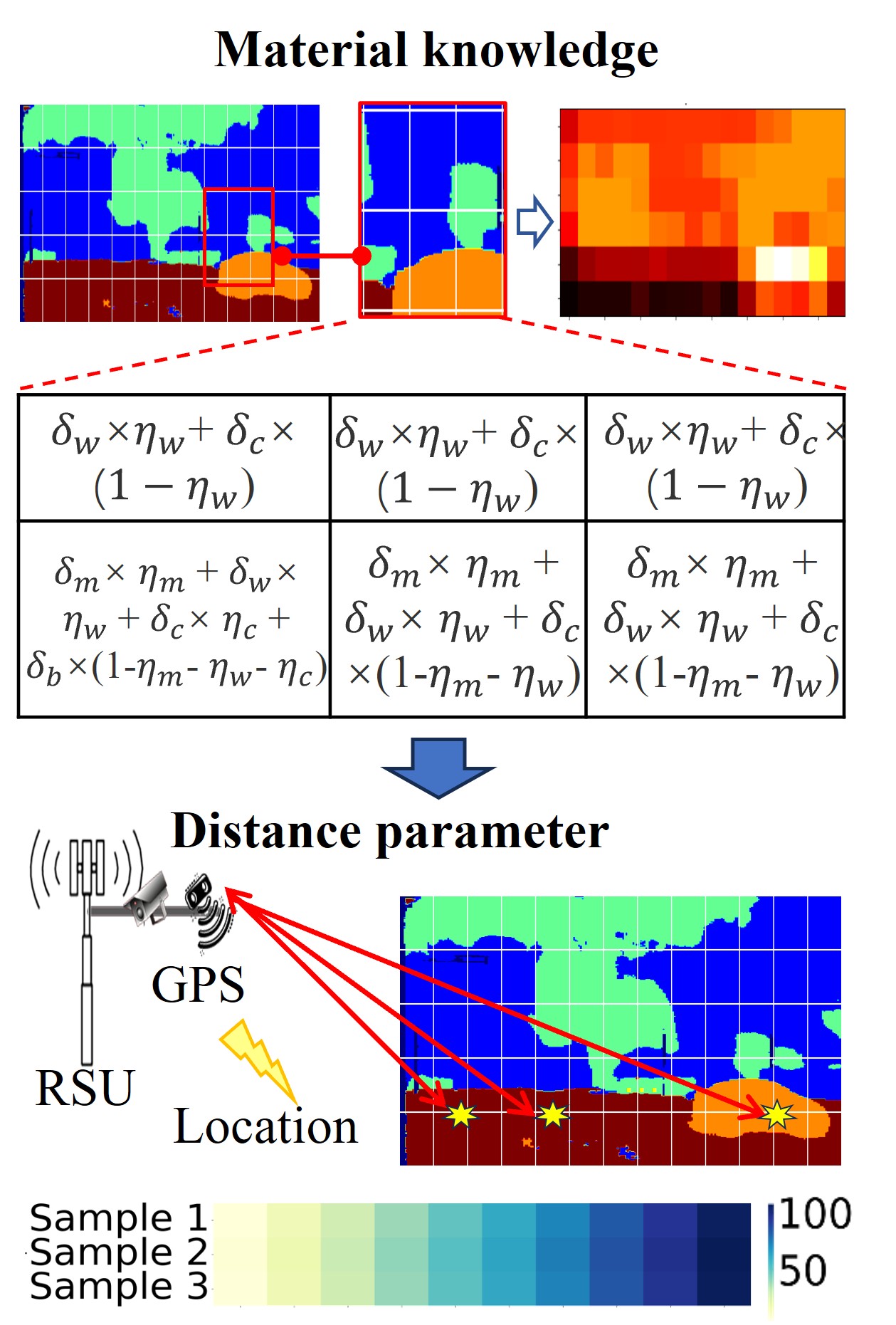}
	\caption{The proposed material and location-aware WEK construction method based on secondary image segmentation results.}\label{REKpre}
\end{figure}

After completing the propagation environment semantic extraction based on segmented images, to further interpret the images from the perspective of EM wave propagation, this paper proposes the material and location-aware WEK construction method based on secondary image segmentation results, as shown in Fig. \ref{REKpre}. The method aims to partition the semantically extracted images into multiple blocks, calculate the reflection characteristics of different materials within each block, and integrate distance features in the propagation environment to construct a relationship between the propagation environment and radio propagation processes that reflects material distribution and distance characteristics. This interpretable and inferable relationship helps improve the accuracy of communication decisions. Meanwhile, compared to directly inputting images containing massive environmental redundancy or image semantics obtained through image processing, directly inputting this relationship can reduce the complexity of network training and time, achieve faster training, and support 6G-oriented online applications, even meeting future requirements for online learning.

The semantically extracted image $ T^{\text{material}}_{h,w} $ is divided into $ N_y \times N_x $ blocks, each with a size of $ B \times B $, where $ B $ is the block side length. The number of blocks is calculated as:
\begin{equation}
N_y = \frac{H}{B} \quad , \quad N_x = \frac{W}{B}.
\end{equation}
In wireless communications, the propagation characteristics of EM waves in media are one of the key factors affecting communication performance \cite{sarkar2003survey}. When EM waves propagate from free space to different media, their propagation behavior undergoes significant changes, mainly reflected in propagation speed, attenuation, reflection, and refraction. The 3GPP TR 38.901 standard provides detailed models and parameters describing the reflection characteristics of EM waves on various material surfaces, including formulas for calculating reflection coefficients under horizontal and vertical polarizations \cite{ETSI_TR_138_901_2018}. For horizontal polarization, the reflection coefficient is calculated as:
\begin{equation}
R_{i}^{T} = \frac{\varepsilon_{r}}{\varepsilon_{0}} \cos (\theta_{T,ZOD}) + \sqrt{\frac{\varepsilon_{r}}{\varepsilon_{0}} - \sin^{2} (\theta_{T,ZOD})}.
\end{equation}
For vertical polarization:
\begin{equation}
R_{L}^{T} = \frac{\cos (\theta_{T,ZOD}) + \sqrt{\frac{\varepsilon_{r}}{\varepsilon_{0}} - \sin^{2} (\theta_{T,ZOD})}}{\cos (\theta_{T,ZOD}) - \sqrt{\frac{\varepsilon_{r}}{\varepsilon_{0}} - \sin^{2} (\theta_{T,ZOD})}}.
\end{equation}

Here, $ \varepsilon_{r} $ represents the complex relative permittivity of various materials, and $ T = \{M(l_{\text{bui}}), M(l_{\text{car}}), M(l_{\text{veg}}), M(l_{\text{go}})\} $, corresponding to four material classifications: cement, metal, wood, and asphalt. The dielectric constants of the materials involved in this work under applicable temperatures are listed in Table \ref{tab5-1}.

\begin{table}[htbp]
\setlength{\tabcolsep}{18pt}
\renewcommand\arraystretch{1.5}
\caption{Reference table for permittivity of different materials at applicable temperatures}\label{tab5-1}
\begin{center}
\begin{tabular}{ccc}
\toprule[1pt]
 {Material}  & {Temperature($^\circ C$)} & {Permittivity}\\
\hline
 {Cement} & {/} & {1.5-2.1}\\
 {Chrome}  & {/} & {7.7-8.0}\\
 {Wood, dry} & {/} & {2/6}\\
 {Asphalt} & {24} & {2.6}\\
\toprule[1pt]
\end{tabular}
\end{center}
\end{table}

For each block $ B_{ij} $, the pixel count of each material label is calculated. Let $ C_{ij}(L_k) $ denote the pixel count of label $ L_k $ in block $ B_{ij} $:
\begin{equation}
C_{ij}(L_k) = \sum_{(x,y) \in B_{ij}} \delta (I(x,y), L_k)\label{5-20},
\end{equation}
where $ k $ is the index of different material labels, and $ \delta(a,b) $ is an indicator function that equals 1 when $ a = b $ and 0 otherwise. The value $ V_{ij} $ of the current block is calculated based on the proportion of each label and its corresponding reflection coefficient. Let $ T_{ij} $ be the total number of pixels in block $ B_{ij} $:
\begin{equation}
V_{ij} = \sum_{k} \left( \frac{C_{ij}(L_k)}{T_{ij}} \cdot \varepsilon_{k} \right).
\end{equation}

\begin{figure*}[!t]
	\centering
	\includegraphics[scale=0.45]{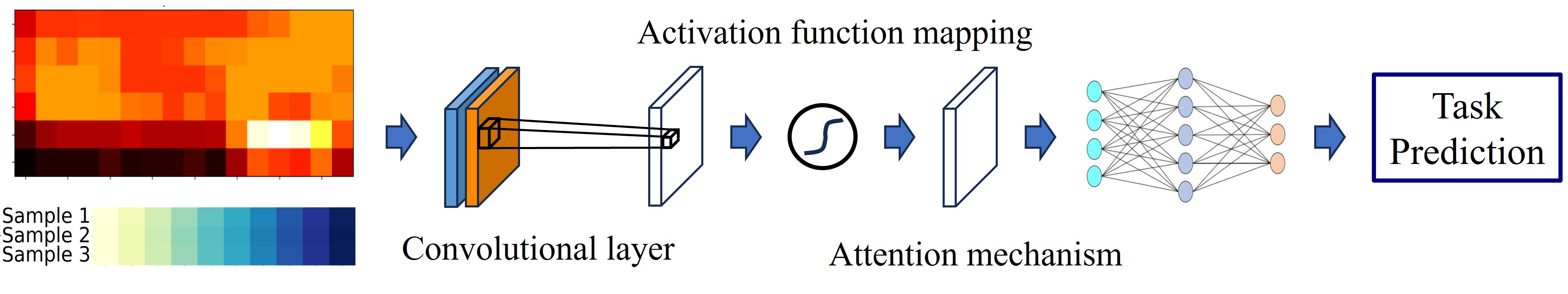}
	\caption{Lightweight CNN combined with an attention mechanism for beam prediction.}\label{CNNpic}
\end{figure*}

For pixels not belonging to known labels, let their count be $ C_{ij}(\text{other}) $, and generate a random value $ r \sim U(0,0.1) $. The contribution to the block value is:
\begin{equation}
V_{ij} = V_{ij} + \left( \frac{C_{ij}(\text{other})}{T_{ij}} \cdot r \right).
\end{equation}

Through the above calculations, the value $ V_{ij} $ for each block is obtained, reflecting the reflection characteristics of different materials within the block. Ultimately, all block values form a material knowledge matrix of size $ N_x \times N_y $, representing the reflection feature distribution across the entire image.

Wireless signal propagation is subject to various attenuation mechanisms, including free-space loss, multipath fading, and environmental obstructions, all of which are strongly influenced by the distance between Tx and Rx \cite{sun2016investigation,phillips2012survey}. According to the free-space model, path loss increases logarithmically with distance, rising by approximately 6 dB for every distance doubling. Real-world effects such as reflection, scattering, and atmospheric absorption further degrade signal quality. Therefore, integrating distance information is essential for accurate propagation prediction and optimization.
In IoV scenarios, Tx and Rx locations are typically acquired via GPS from base stations, RSUs, or mobile terminals, with minimal privacy or security overhead \cite{lu2018survey}. To utilize this geographic data, latitude and longitude coordinates are converted into physical distance using the Haversine formula, which accounts for Earth's curvature and provides an accurate metric for wireless analysis.
Given two points with coordinates $ (\text{lat}_1, \text{lon}_1) $ and $ (\text{lat}_2, \text{lon}_2) $, their longitude difference $ \Delta \text{lon} $ and latitude difference $ \Delta \text{lat} $ are calculated after converting degrees to radians. Using the Haversine formula \cite{winarno2017location} and Earth’s average radius $ r $, the actual distance $ d $ is computed as:
\begin{equation}
d = r \cdot c,
\end{equation}
\begin{equation}
c = 2 \cdot \arcsin (\sqrt{a}),
\end{equation}
\begin{equation}
a = \sin^2 \left( \frac{\Delta \text{lat}}{2} \right) + \cos (\text{lat}_1) \cdot \cos (\text{lat}_2) \cdot \sin^2 \left( \frac{\Delta \text{lon}}{2} \right).\label{eq20}
\end{equation}

The resulting distance $ d $ serves as a geometric distance feature, effectively characterizing the relationship between environmental information and radio propagation. Together with $ V_{ij} $, it forms the material- and location-related wireless environment knowledge, providing input for subsequent decision network modules. The effectiveness of this knowledge is validated through beam prediction tasks.

\subsection{Lightweight Convolutional Attention Prediction Network}

\begin{figure*}[htp]
  \centering
    \subfigure[Single vehicle driving]{\includegraphics[width=0.3\textwidth]{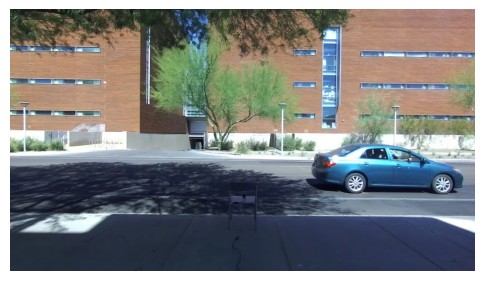}}
    \qquad
    \subfigure[Same type of vehicle driving]
    {\includegraphics[width=0.3\textwidth]{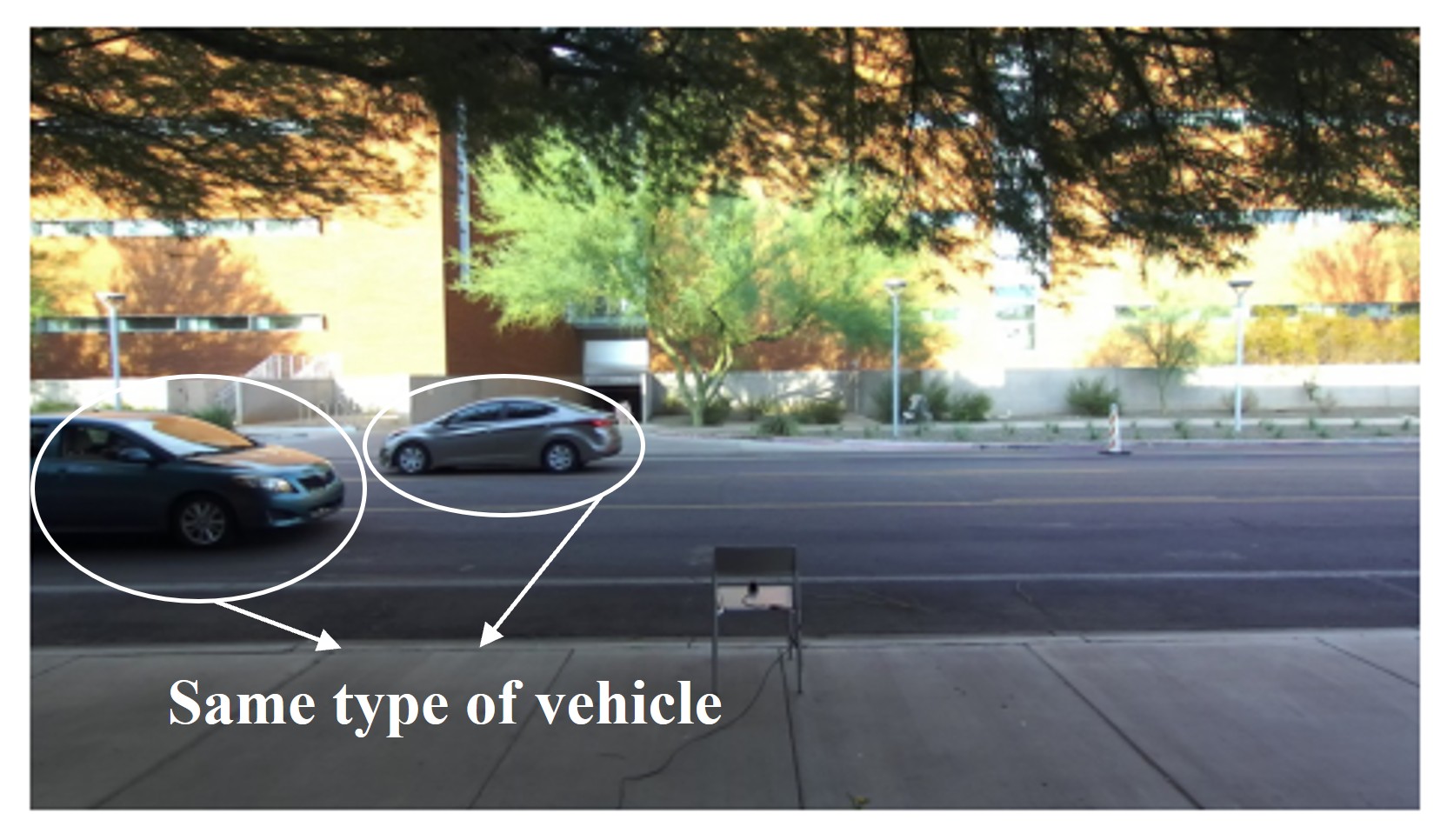}}\\
    \subfigure[Multiple types of vehicles driving]{\includegraphics[width=0.3\textwidth]{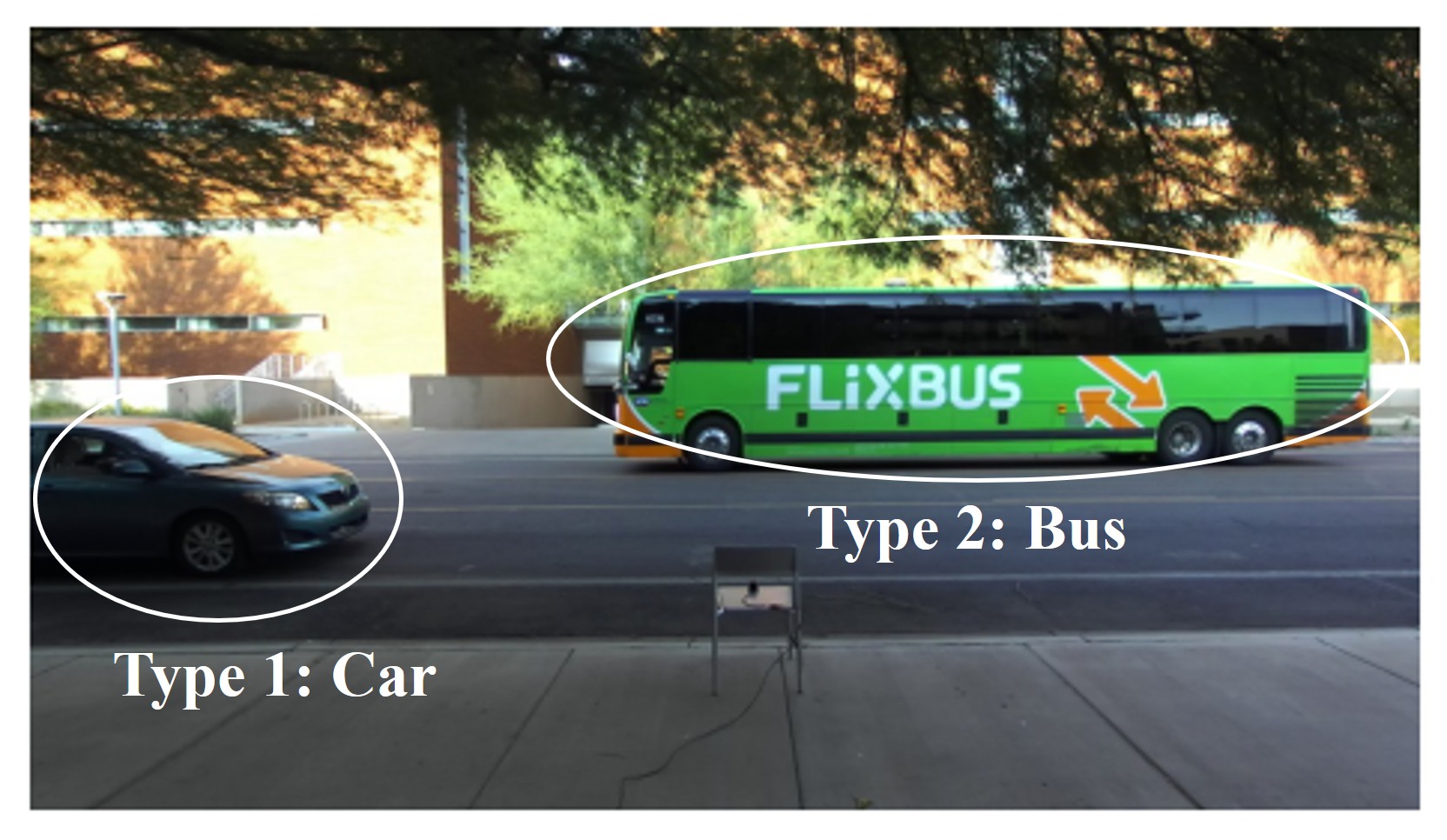}}
    \qquad
    \subfigure[Multiple types of vehicles and pedestrians]
    {\includegraphics[width=0.3\textwidth]{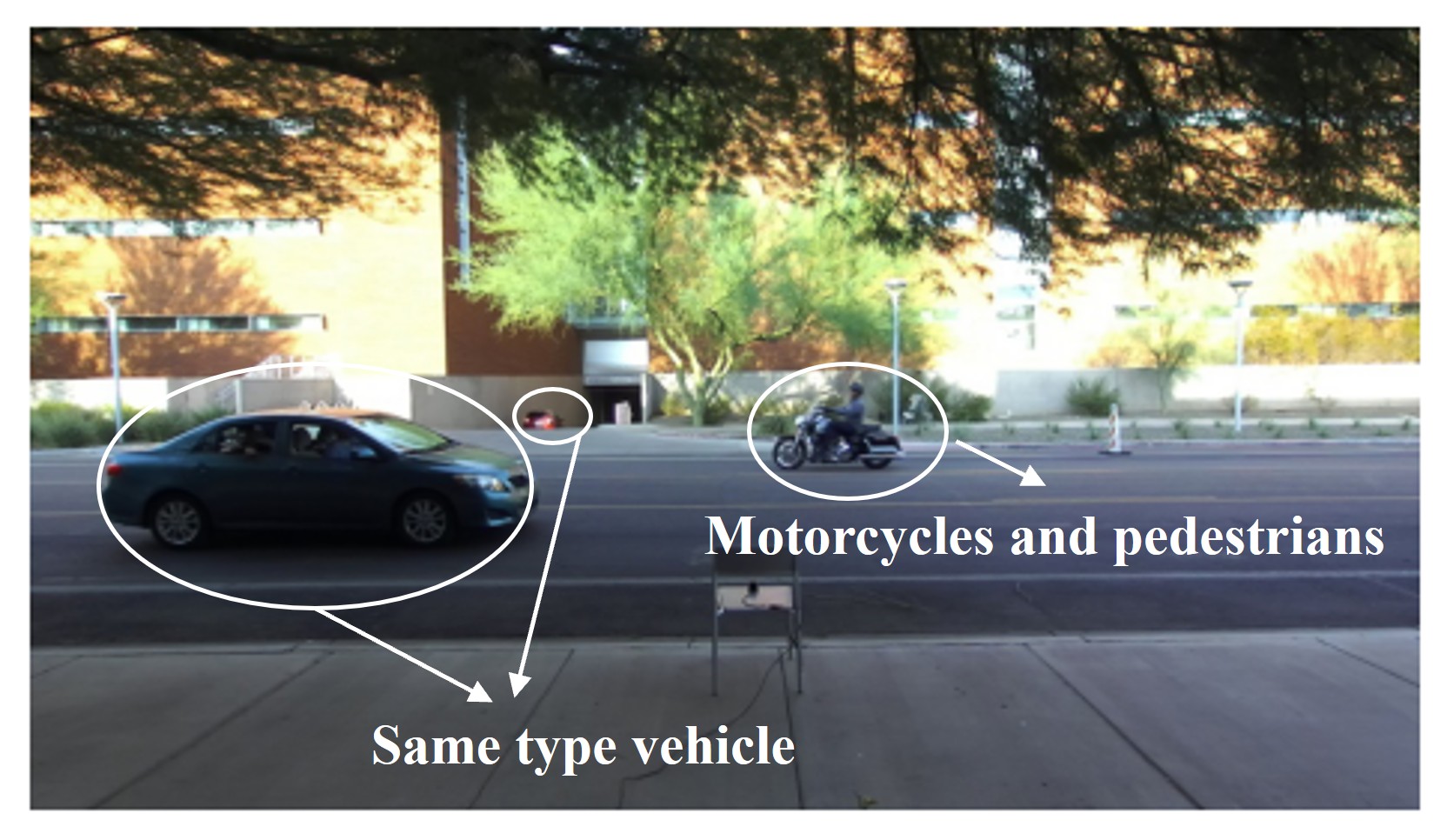}}\\    
\caption{Four different types of random image samples from the DeepSense 6G dataset}
\label{datasetIntro}
\end{figure*}

The decision network module adopts a lightweight convolutional neural network (CNN) combined with an attention mechanism, as shown in Fig \ref{CNNpic}. CNN excels at processing structured data, particularly in image processing tasks. It extracts local features through convolutional layers and progressively learns high-level feature representations of data using multiple convolutional layers and nonlinear activation functions. The attention mechanism dynamically allocates weights to different parts of the input data, enabling the network to focus on more task-relevant features, thereby enhancing model performance and generalization capabilities\cite{tian2020attention}. The input to the beam prediction module can be expressed as:
\begin{equation}
x_{bp} = \left[ (V_{ij})_{N_y \times N_x}, (d)_{1 \times 1}, (y_{bp, lab})_{1 \times 1} \right],
\end{equation}
where $ y_{\text{target}} $ represents the pre-labeled beam target at the current Rx position for supervised learning. The $ N_y \times N_x $-dimensional material knowledge matrix in the input data reflects the reflection characteristics of diverse materials, characterizing the influence of scatterers with varying material properties on radio wave propagation. Despite the small data dimensions and high similarity, numerical variations significantly describe radio wave propagation processes in complex environments. Given these input characteristics, the lightweight CNN with attention offers two advantages: (1) The small input dimensions reduce computational complexity and parameter count. A few convolutional layers directly extract local features without complex pooling operations, preserving more detailed information. (2) The attention mechanism dynamically assigns weights to highlight features critical to beam prediction, improving the model’s sensitivity to subtle variations in highly similar data.

The lightweight CNN extracts features through $ L $ convolutional layers, each containing $ C_L $ convolutional kernels of size $ K_L \times K_I $ with a stride of 1 (padding=$\frac{K_I - 1}{2}$). Each convolutional layer is followed by a ReLU activation function, producing an output of size $ N_y \times N_x \times C_L $, where $ C_L $ is the number of channels in the final convolutional layer. The output of the $ l $-th convolutional layer is:
\begin{equation}
X^{(l)} = \text{ReLU} \left( W^{(l)} * X^{(l-1)} + b^{(l)} \right),
\end{equation}
where $ W^{(l)} $ and $ b^{(l)} $ are the convolutional kernel and bias of the $ l $-th layer, respectively, and $ X^{(l)} $ is the input material knowledge matrix. The CNN output $ X^L $ is flattened into a 2D tensor $ X_{\text{flat}} \in \mathbb{R}^{(N_y \times N_x) \times C_L} $. While these features may contain abundant information, not all are equally important for beam prediction. The attention mechanism dynamically assigns a weight to each feature, reflecting its importance to the task. A “context vector” is generated by weighted summation of all features, capturing the most relevant information:
\begin{equation}
\text{context} = \left( \text{softmax} \left( \mathbf{V}_a \left( \tanh \left( \mathbf{W}_a X_{\text{flat}} + \mathbf{U}_a X_{\text{flat}} \right) \right) \right) \right)^T X_{\text{flat}},
\end{equation}
where $ \mathbf{V}_a $, $ \mathbf{W}_a $, and $ \mathbf{U}_a $ are weight matrices of the attention module, $ \alpha $ represents attention weights, and “context” is the weighted context vector. The distance $ d $ is linearly transformed and concatenated with the context vector:
\begin{equation}
X_{\text{final}} = \text{concat} \left( \text{context}, d \right).
\end{equation}
Finally, the beam prediction result is output through a fully connected (FC) layer:
\begin{equation}
y_{bp} = f_{Lbp} \left( X_{\text{final}} \right). 
\end{equation}

The loss function for the beam prediction module uses cross-entropy loss:
\begin{equation}
\mathcal{L}_{bp} = -\sum_{i=1}^{N} \log \left( \frac{\exp \left( y_{bp} \left[ y_{bp,lab}^{(i)} \right] \right)}{\sum_{m=1}^{M_{bp}} \exp \left( y_{bp} \left[m\right] \right)} \right),
\end{equation}
where $ N $ is the batch size, $ M_{bp} $ is the total number of beam classes, and $ y_{bp,lab}^{(i)} $ is the true beam label of the $ i $-th sample. The prediction accuracy is evaluated using the Top-K accuracy metric:
\begin{equation}
A_K = \frac{1}{N_{\text{tot}}} \sum_{n=1}^{N_{\text{tot}}} \left( y_{bp,lab}^{(n)} \in \text{Top-}K \left( y_{bp}^{(n)} \right) \right) \times 100\% ,
\end{equation}
where $ N_{\text{tot}} $ is the total number of test samples, $ \text{Top-}K \left( y_{bp}^{(n)} \right) $ denotes the top $ K $ predicted beam classes for the $ n $-th sample, and $ \mathbb{I}(\cdot) $ is an indicator function checking whether the true label is among the top $ K $ predictions.

Additionally, this paper proposes a novel PCEI to comprehensively evaluate prediction accuracy and training time. The index combines model accuracy with a logarithmic transformation of training time and introduces a time stability penalty, offering a more balanced and fair evaluation perspective. Its formula is:
\begin{equation}
\text{PCEI} = \frac{P}{\log (T+1)} \cdot e^{-\alpha \cdot \sigma_T},
\end{equation}
\begin{equation}
\sigma_T = \sqrt{\frac{1}{n} \sum_{i=1}^{n} \left( t_i - \mu(T) \right)^2}, 
\end{equation}
where $ P $ is prediction accuracy (model performance), $ T $ is training time, $ \alpha $ is a time penalty coefficient adjusting the impact of training time fluctuations, and $ \sigma_T $ is the coefficient of variation of training time, indicating relative fluctuations. A higher $ \sigma_T $ implies greater instability in training time. The logarithmic term $ \log(T+1) $ mitigates the linear impact of training time growth, preventing excessive penalty for longer times. The exponential decay term $ e^{-\alpha \cdot \sigma_T} $ penalizes time instability, encouraging consistent and stable training processes.

Compared to traditional methods that separately evaluate accuracy and computational complexity, PCEI holistically considers prediction accuracy and training time stability, avoiding biases from simplistic linear relationships. By incorporating logarithmic scaling and exponential decay, PCEI captures the nonlinear relationship between performance gains and training time growth, reduces uncertainty in training time, and more accurately reflects real-world performance variations, making it suitable for high-reliability practical applications.

\begin{table*}[htb]
\setlength{\tabcolsep}{8pt}
\renewcommand\arraystretch{1.3}
\caption{Network architecture and parameter setting of PES-WEKBP and comparison methods}\label{tab5-2}
\begin{center}
\begin{tabular}{c|c|c|c|c|c|c|c|c}
\toprule[1pt]
  Module  & \multicolumn{4}{c|}{PES-WEKBP} & \multicolumn{4}{c}{Contrast method} \\ 
\hline
  \makecell{Input size} &  {\makecell{$B = 20$:\\$27 \times 48$}}  & {\makecell{$B = 30$:\\$18 \times 32$}}
  & {\makecell{$B = 40$:\\$13 \times 24$}}
  & {\makecell{$B = 50$:\\$10 \times 19$}}
  & $256 \times 256$  & $128 \times 128$  & $64 \times 64$  & $32 \times 32$ \\
\hline

  \makecell{Image semantics\\segmentation module}   &   \multicolumn{8}{c}{\makecell{Encoder: (OPE-ESA-MixFFN)$\times $4\\Decoder: MLP-Dropout-Conv}} \\
\hline

    \makecell{Beam prediction\\module} & \multicolumn{4}{c|}{\makecell{Conv-BN-ReLu\\Conv-BN-ReLu\\FC-BN-ReLu\\Dropout\\FC-BN-ReLu\\Attention\\FC}}  &  \multicolumn{2}{c|}{\makecell{Conv-BN-ReLu\\MaxPool\\Conv-BN-ReLu\\MaxPool\\FC-BN-ReLu\\Dropout\\FC-BN-ReLu\\Attention\\FC}}  &  \multicolumn{2}{c}{\makecell{Conv-BN-ReLu\\Conv-BN-ReLu\\FC-BN-ReLu\\Dropout\\FC-BN-ReLu\\Attention\\FC}} \\
\hline

  Kernal size &  \multicolumn{4}{c|}{3, 3} & \multicolumn{4}{c}{3, 3} \\ 
\hline

  Stride & \multicolumn{4}{c|}{/} & \multicolumn{4}{c}{4, 4} \\ 
\hline

  Batch size & 32 & 32 & 32 & 64 & 128 & 128 & 64 & 32 \\
\hline

  Learning rate & 0.001 & 0.0005 & 0.001 & 0.001 & 0.001 & 0.0005 & 0.001 & 0.0025 \\
\hline

 Feature fusion& \multicolumn{4}{c|}{linear transformation splicing} & \multicolumn{4}{c}{\makecell{normalized linear transform \\ features fusion}}\\ 
\toprule[1pt]
\end{tabular}
\end{center}
\end{table*}

\section{Simulation verification and result analysis}\label{FOUR}

\subsection{Dataset Introduction}

The DeepSense 6G Scenario 9 dataset validates beam prediction performance within a simulated V2I mmWave communication scenario \cite{alkhateeb2023deepsense}. This dataset contains 5,964 multi-modal samples. 
Environmental sensing data is provided by RGB images with a resolution of $960 \times 540$ pixels at 30 frames per second. High-precision spatiotemporal alignment relies on GPS-RTK coordinates.
The mmWave signal characteristics are captured by 64-dimensional received power vectors derived from a 16-element 60 GHz phased array. 
The experimental setup consists of a fixed roadside unit equipped with an RGB camera, LiDAR, and a mmWave transceiver, along with a mobile vehicle unit containing a mmWave transmitter and GPS-RTK.
Data collection was conducted on a bidirectional street 10.6 meters wide, featuring mixed traffic including cars, trucks, and buses operating under a 25 mph speed limit. Scenarios such as multi-vehicle occlusion and bidirectional movement are captured, as shown in Fig. \ref{datasetIntro}.

For model training, the dataset is split into 80\% training and 20\% test sets.  A custom data loader efficiently processes batches by encapsulating dataset loading, preprocessing, and tensor conversion logic.
RGB images, positional coordinates, and beam labels are indexed and retrieved to exploit multi-modal correlations.
This structured preprocessing ensures compatibility with neural network training while preserving the dataset’s spatiotemporal and signal diversity, establishing a rigorous foundation for beam prediction evaluation.

\subsection{Comparative Methods and Network Parameter Settings}
Existing studies have demonstrated that leveraging deep neural networks and environmental information to directly utilize image semantic segmentation results can achieve relatively ideal outcomes \cite{yang2023environment}.
In contrast, the proposed PES-WEKBP method establishes a quantifiable relationship between environmental features and wireless propagation processes, inputting weighted wireless propagation knowledge graphs to significantly reduce neural network training time complexity while maintaining accuracy.

In this study, the comparative method directly uses image semantic segmentation results for beam prediction. The proposed PES-WEKBP method and the comparative method share identical operations in the image semantic segmentation module, with the primary difference lying in the data input to the decision network module. The PES-WEKBP method inputs distance features extracted from the environment and material and location-aware WEK graphs, incorporating environmental information such as vehicles, sidewalks, roads, shadows, vegetation, buildings, and lampposts. In contrast, the comparative method inputs user location information and image semantic segmentation results into the decision network module, referencing conclusions on effective environmental semantics from literature\cite{yang2023environment}, and considers fewer environmental elements, including only vehicles, sidewalks, and roads. To ensure fairness, identical network structures are adopted in the decision network module to validate the effectiveness of the proposed knowledge representation. Based on the characteristics of the input data for both methods, network parameters are adaptively matched through neural architecture search to ensure optimal feature learning. Specific network architectures and parameter settings are detailed in Table \ref{tab5-2}.

\begin{table*}[htb]
\setlength{\tabcolsep}{9pt}
\renewcommand\arraystretch{1.5}
\caption{Four samples under the category of ${\rm{EC}}{{\rm{R}}^3}$}\label{tab5-3}
\begin{center}
\begin{tabular}{cccc}
\toprule[1pt]
 {Random samples category} & {Number of image segmentation} & {Number of PES exaction} & {${\rm{EC}}{{\rm{R}}^3}$}\\
\hline
 {Single vehicle driving} & {10} &{4} & {60\%}\\
 {Same type of vehicle driving} & {11} & {4} & {63.6\%}\\
 {Multiple types of vehicles driving} & {11} & {4} & {63.6\%}\\
 {Multiple types of vehicles and pedestrians} & {12} & {4} & {66.7\%}\\
\toprule[1pt]
\end{tabular}
\end{center}
\end{table*}

\begin{figure*}[htb]
	\centering
	\includegraphics[scale=0.3]{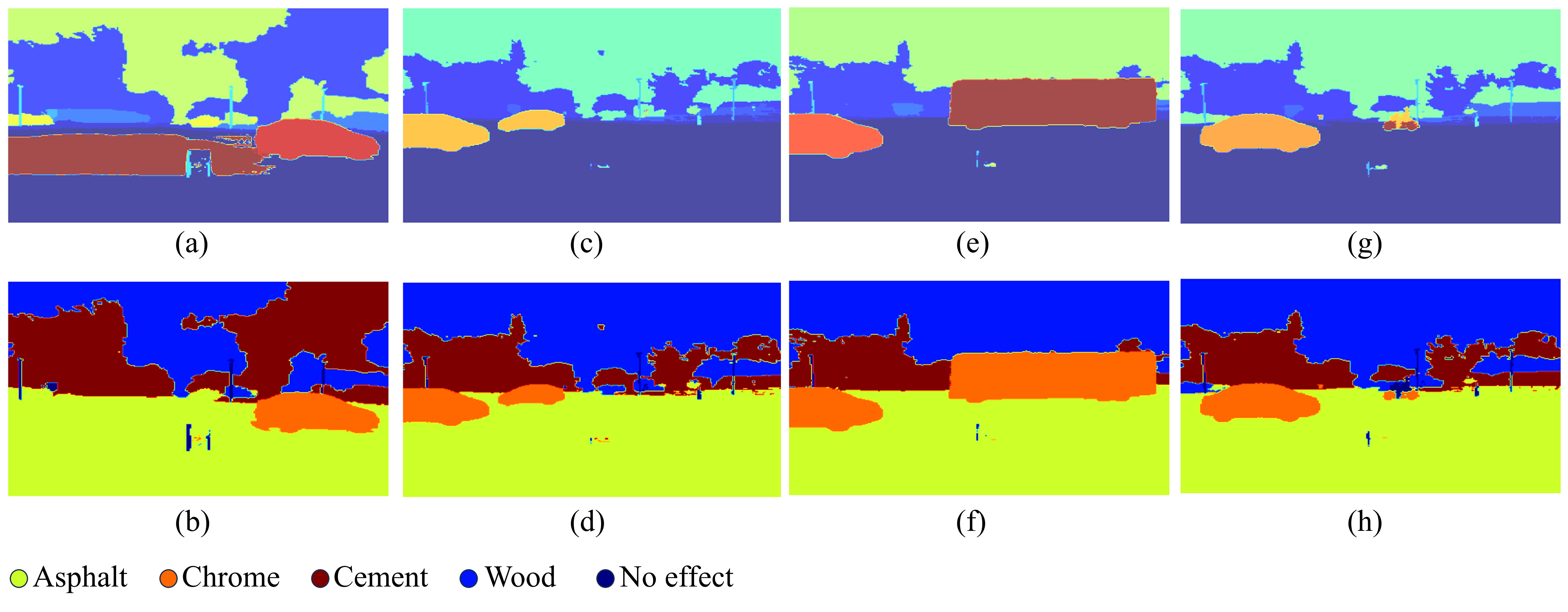}
	\caption{Image segmentation and PES extraction results.}\label{REK-All}
\end{figure*}

In the image semantic segmentation module, both methods use a pre-trained Segformer network without retraining. For the decision network module, the input dimensions of the PES-WEKBP method correspond to the knowledge graph sizes generated during block partitioning with varying side lengths $ B = \{20, 30, 40, 50\} $. 
The comparative method adopts common image processing input sizes. 
Since large-scale images ($256 \times 256 $ and $128 \times 128$) cannot be effectively processed by the lightweight CNN, the network architecture is optimized by adding two pooling layers to minimize complexity. The symbol ``-" denotes sequential stacking, ``$\left(  \cdot  \right) \times 4$" indicates four repetitions of internal layers.

\subsection{Simulation Results of Knowledge Representation Based on Propagation Environment Semantics}
In this section, we validate the knowledge representation results based on propagation environment semantics through simulations. To evaluate the effectiveness of the proposed PES extraction method using image segmentation, the environmental category redundancy reduction rate ($\text{ECR}^3$) metric is introduced. $\text{ECR}^3$ measures the reduction in redundant image categories during processing, reflecting the simplification degree of semantic information in images. Its calculation formula is:
\begin{equation}
\text{ECR}^3 = \left(1 - \frac{C_p}{C_o}\right) \times 100\%,
\end{equation}
where $ C_p $ denotes the number of PES-extracted categories, and $ C_o $ represents the number of original image segmentation categories.
Table \ref{tab5-3} lists the $\text{ECR}^3$ values under different traffic scenarios, demonstrating the redundancy reduction effectiveness of the proposed method.

Fig. \ref{REK-All} illustrates the image segmentation and PES extraction results across various driving scenarios. Overall, the method effectively reduces environmental complexity by merging propagation-relevant objects into key material categories, achieving a significant $\text{ECR}^3$ in each case.
For instance, in the single-vehicle scenario (Fig. \ref{REK-All}(a)(b)), segmentation identifies multiple objects, while PES retains critical materials such as asphalt—combining roads and shadows—and filters out non-impact items like lampposts. This reduces 10 categories to 4, with an $\text{ECR}^3$ of 60\%, confirming accurate environmental simplification for beam prediction.
In multi-vehicle scenarios (Fig. \ref{REK-All}(c)(d) and (e)(f)), the method consolidates cars, buses, and vegetation into metal and other material attributes, ignoring irrelevant distinctions. Both cases achieve an $\text{ECR}^3$ of 63.6\%.
In a complex scene with pedestrians (Fig. \ref{REK-All}(g)(h)), cars and motorcycles are merged into metal, while pedestrians are removed due to minimal propagation impact. This results in a reduction from 12 to 4 categories and an $\text{ECR}^3$ of 66.7\%, validating the method’s efficacy in complex traffic environments.

\begin{figure*}[!t]
	\centering
	\includegraphics[scale=0.22]{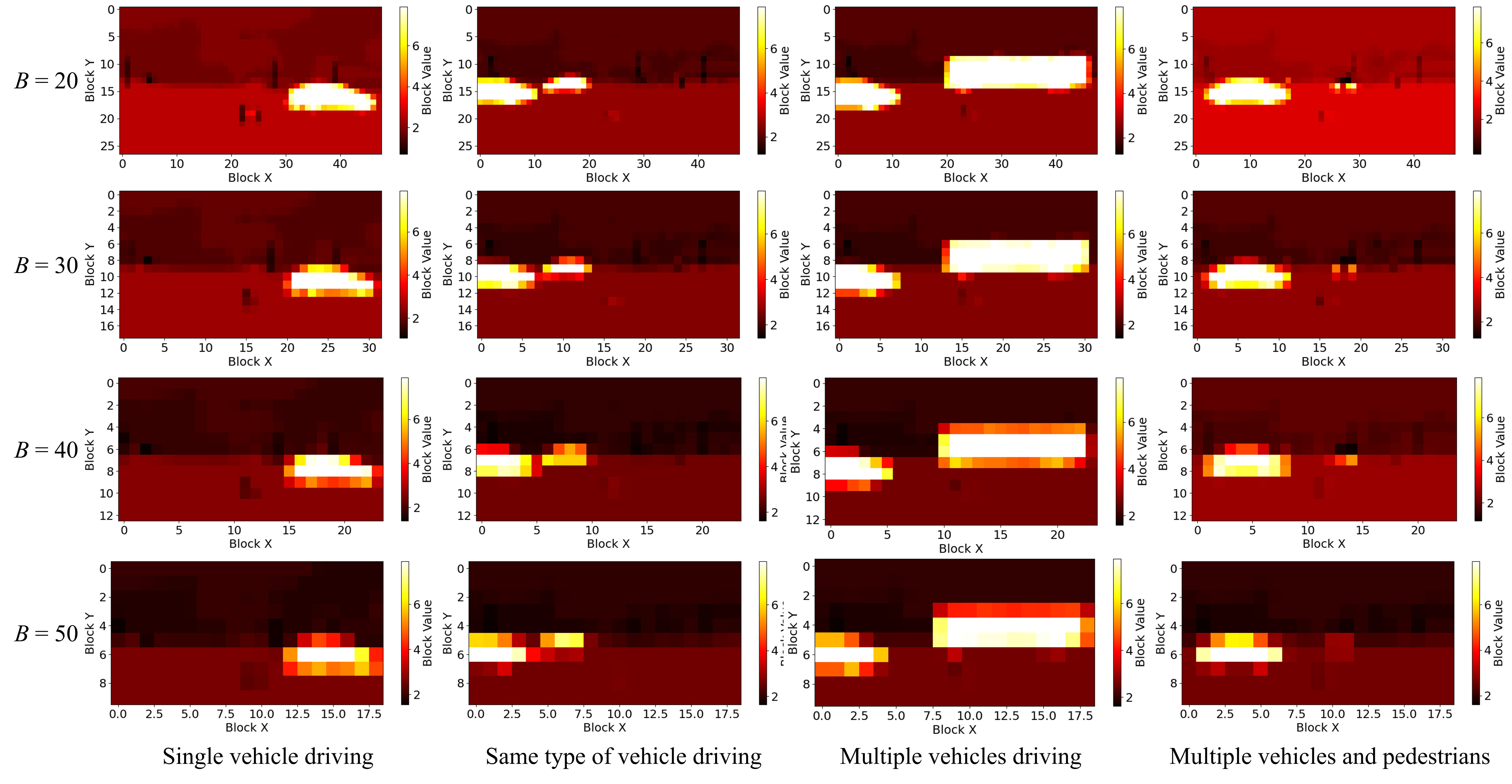}
	\caption{Image segmentation and PES extraction results of image segmentation and PES extraction results.}\label{REKgraph-All}
\end{figure*}

To quantitatively assess the reduction of information redundancy during propagation environment semantic extraction, the environmental information redundancy reduction rate ($\text{EIR}^3$) is proposed, defined as:
\begin{equation}
\text{EIR}^3 = \left(1 - \frac{I_o}{I_p}\right) \times 100\%,
\end{equation}
where $ I_p $ is the total pixel count of processed images, and $ I_o $ is the total pixel count after PES extraction. This metric evaluates redundancy compression from a data-dimensional perspective, with higher values indicating stronger redundancy elimination.

\begin{table}[htbp]
\setlength{\tabcolsep}{8.5pt}
\renewcommand\arraystretch{1.5}
\caption{Reduction rate of environmental information redundancy under the partition size of four kinds of knowledge graph blocks}\label{tab4}
\begin{center}
\begin{tabular}{ccc}
\toprule[1pt]
 {Square partition size} & {Size of knowledge graph} & {{$\text{EIR}^3$}} \\
\hline
 {$B = 20$} & {$27 \times 48$} &{99.75\%} \\
 {$B = 30$} & {$18 \times 32$} & {99.89\%} \\
 {$B = 40$} & {$13 \times 24$} & {99.94\%} \\
 {$B = 50$} & {$10 \times 19$} & {99.96\%} \\
\toprule[1pt]
\end{tabular}
\end{center}
\end{table}

Table \ref{tab4} lists the $\text{EIR}^3$ values under different block partitioning sizes. As the block side length $ B $ increases from 20 to 50, $\text{EIR}^3$ improves from 99.75\% to 99.96\%, indicating exponential data compression. Despite drastic dimensionality reduction, the material and location-aware WEK representation preserves critical EM propagation features through weighted material reflection coefficients and distance metrics. For example, using the Haversine formula, Tx-Rx distance $ d $ combined with material distribution generates a comprehensive knowledge spectrum that distinguishes near-field strong reflection zones. 
Even at $ B = 50 $, material and location-aware WEK retains spatial distributions and influence intensities of key scatterers, transforming unweighted pixels into weighted knowledge spectra.

\begin{figure*}[!htb]
  \centering
    \subfigure[Top-1]{\includegraphics[width=0.38\textwidth]{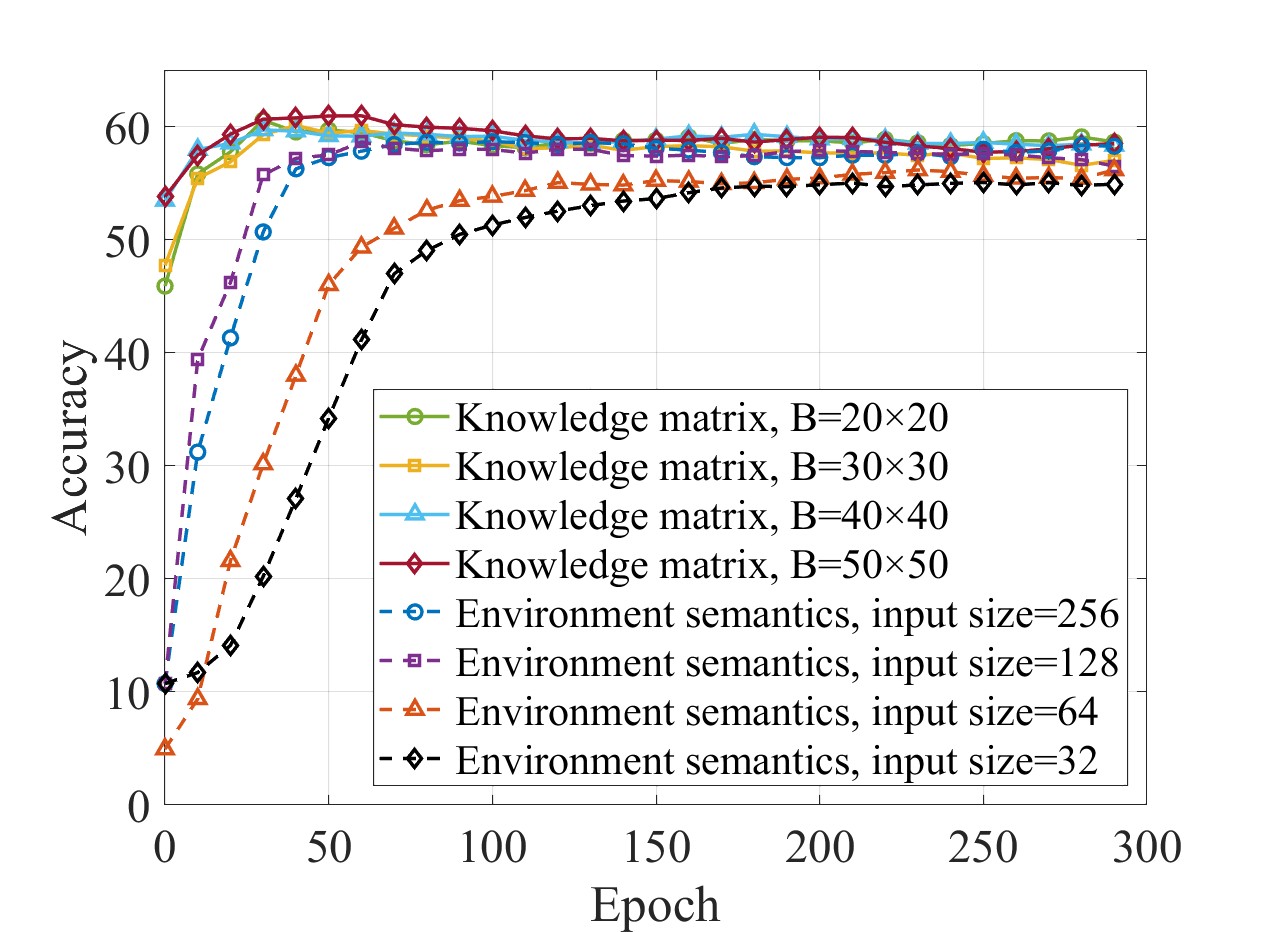}}
    \qquad \qquad
    \subfigure[Top-2]
    {\includegraphics[width=0.38\textwidth]{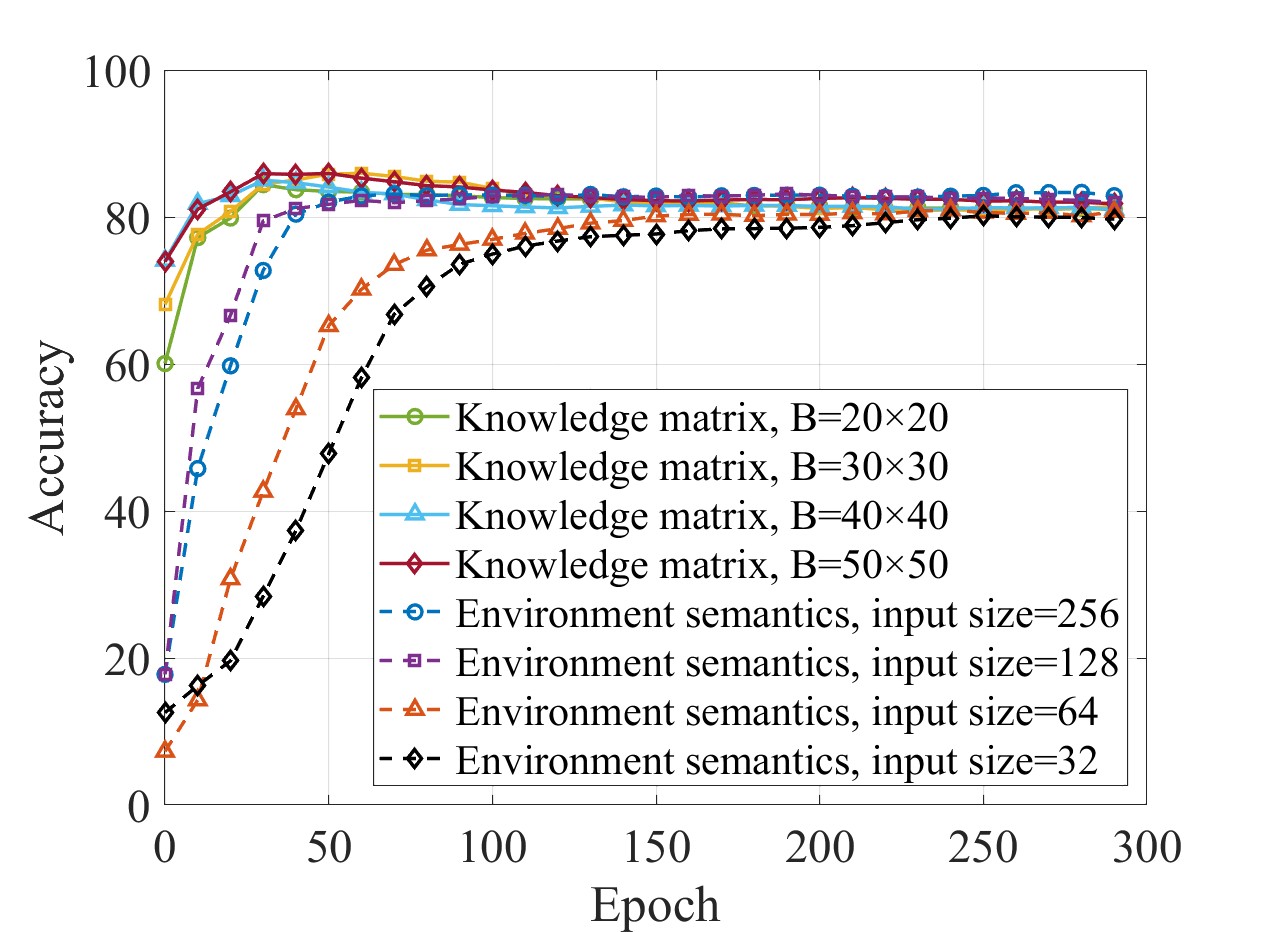}} \\
    \subfigure[Top-3]{\includegraphics[width=0.38\textwidth]{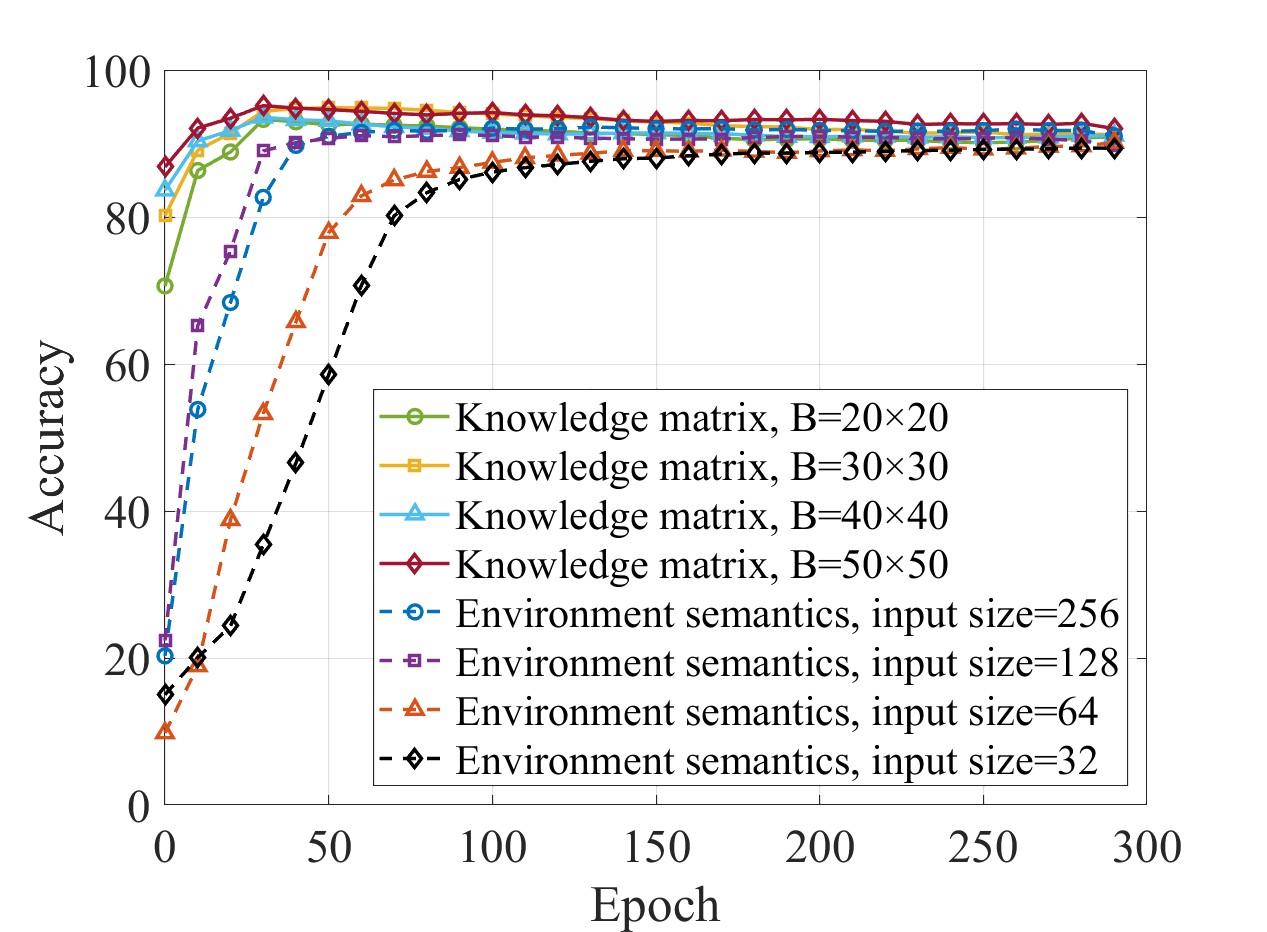}}
    \qquad \qquad
    \subfigure[Top-4]
    {\includegraphics[width=0.38\textwidth]{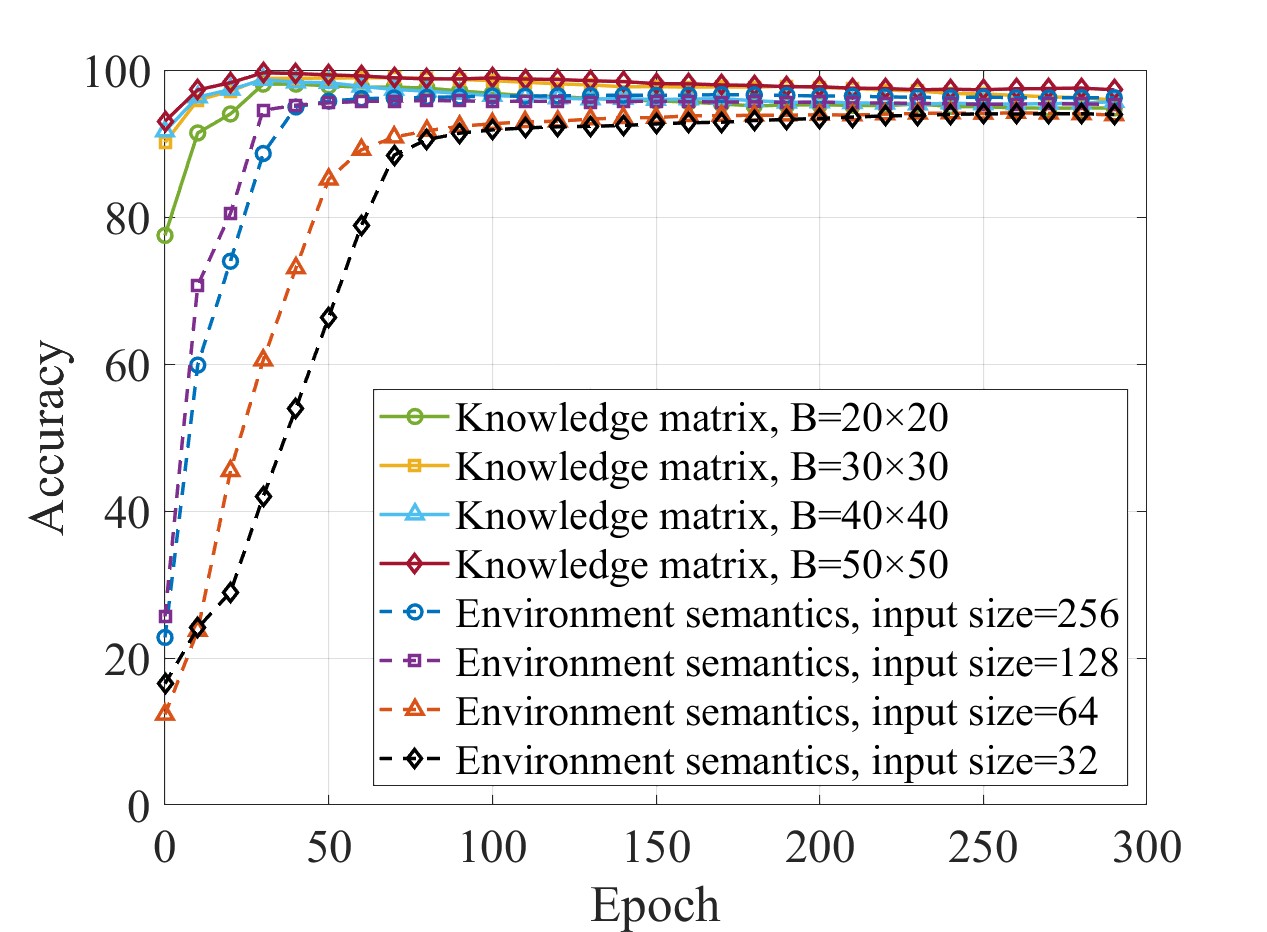}} 
\caption{Comparison curves of PES-WEKBP and Top-G beam prediction accuracy of the comparison method under various input sizes.}
\label{BeamPrePlot}
\end{figure*}

Fig. \ref{REKgraph-All} displays WEK representation results for single-vehicle, homogeneous-vehicle, multi-vehicle, and pedestrian-inclusive scenarios under different block sizes ($ B = 20, 30, 40, 50 $). 
The WEK transformation process converts RGB inputs into material knowledge matrices that quantify EM propagation characteristics. The original 10-12 semantic categories are reduced to four propagation-relevant material classes: cement, metal, wood and asphalt, while omitting non-essential elements like lampposts and pedestrians. 
Based on Equation \ref{5-20}, each block’s value $ V_{ij} $ is calculated via material proportion weighting.
Despite $\text{EIR}^3$ exceeding 99\%, WEK maps effectively distinguish key propagation regions. 
The 20-unit block configuration with $48 \times 27$ resolution preserves fine material variations, whereas the 50-unit blocks at $10 \times 19$ resolution emphasize clustered metallic scatterers. 
In pedestrian scenarios in the last column of Fig. \ref{REKgraph-All}, vehicles retain high weights, vegetation and buildings show intermediate values, and pedestrians are excluded. This ``unweighted-to-weighted" transformation enables both data compression and precise multipath effect modeling for beam prediction.

\subsection{Beam Prediction Simulation Results Based on PES-WEKBP}

\begin{table*}[ht]
\setlength{\tabcolsep}{7pt}
\renewcommand\arraystretch{1.5}
\caption{Comparison table of Top-G beam prediction accuracy of PES-WEKBP and comparison method under different input sizes.}\label{Accuracytab}
\begin{center}
\begin{tabular}{cccccccccc}
\toprule[1pt]
  \multicolumn{2}{c}{} & \multicolumn{4}{c}{\textbf{PES-WEKBP}} & \multicolumn{4}{c}{\textbf{Comparison method}} \\ 
\hline
  \multicolumn{2}{c}{\textbf{Input size}} &  {\makecell{$B = 20$:\\$27 \times 48$}}  & {\makecell{$B = 30$:\\$18 \times 32$}}
  & {\makecell{$B = 40$:\\$13 \times 24$}}
  & {\makecell{$B = 50$:\\$10 \times 19$}}
  & $256 \times 256$  & $128 \times 128$  & $64 \times 64$  & $32 \times 32$ \\

 \multicolumn{2}{c}{\textbf{Iterations number}}  &  50  & 50 & 50 & 70
  & 80  & 80  & 150  & 150 \\
 \multirow{5}*{\makecell{Accuracy\\(\%)}}
     & Top-1 & 61.89 & 61.80 & 60.67 & 63.68 & 59.51 & 59.85 & 56.92 & 55.49 \\
     & Top-2 & 84.61 & 87.20 & 85.90 & 87.24 & 84.07 & 84.41 & 81.56 & 80.55 \\
     & Top-3 & 93.41 & 95.16 & 94.54 & 95.54 & 92.62 & 91.76 &  90.36 & 89.44 \\
     & Top-4 & 98.85 & 99.10 & 99.37 & 99.74 & 96.81 & 96.14 & 94.47 & 94.22 \\
     & Top-5 & 98.94 & 99.91 & 99.48 & 99.92 & 98.58 & 98.24 & 96.90 & 96.81 \\
\toprule[1pt]
\end{tabular}
\end{center}
\end{table*}

This section validates the feasibility and advantages of the proposed PES-WEKBP method through beam prediction tasks. The verification scheme employs material and location-aware WEK graphs based on PES with varying block sizes ($ B = 20, 30, 40, 50 $) as input, while the comparative scheme uses image segmentation-derived environmental semantics with standard image sizes. The comparative method represents state-of-the-art research, directly utilizing high-resolution image semantic segmentation results but without integrating material reflection characteristics or distance features. To ensure fairness, both schemes adopt identical lightweight CNN architectures with attention mechanisms and optimize network parameters via neural architecture search to adapt to different input characteristics.

Fig. \ref{BeamPrePlot} compares the Top-$ G $ beam prediction accuracy curves of the knowledge representation method and the environment semantic-based method under different input sizes. Experimental results demonstrate that the proposed PES-WEKBP method outperforms the baseline across all Top-$ K $ metrics. For example, in Top-1 accuracy, PES-WEKBP achieves over 60\% at $ B = 50 $, whereas the comparative method fails to reach 60\% even with $256 \times 256$ inputs. 
Notably, at the reduced input dimension of 50 units corresponding to a $19 \times 10$ resolution, the proposed method maintains Top-4 accuracy exceeding 99\%,  outperforming the baseline's results achieved with standard $256 \times 256$ pixel inputs.
The proposed method achieves comparable performance to the baseline while reducing input data volume to 0.04\% of original images through 50-unit block compression, demonstrating precise feature extraction with minimal redundancy.

Furthermore, the PES-WEKBP method exhibits significantly faster convergence. As shown in Fig. \ref{BeamPrePlot} and Table \ref{Accuracytab}, for an input size of $ B = 20 $, PES-WEKBP achieves 61.89\% Top-1 accuracy with only 30 training iterations, whereas the baseline requires 50 iterations for 59.51\% accuracy under $256 \times 256$ inputs. In extreme cases, even with 150 iterations, the baseline fails to match the proposed method’s 63.68\% accuracy at $ B = 50 $. This highlights how material and location-aware WEK knowledge graphs reduce model complexity, accelerate convergence, and minimize training resource consumption.

\begin{table*}[ht]
\setlength{\tabcolsep}{9pt}
\renewcommand\arraystretch{1.5}
\caption{Top-1 beam prediction accuracy and complexity balance comparison table of PES-WEKBP and comparison method}\label{tab57}
\begin{center}
\begin{tabular}{ccccccccc}
\toprule[1pt]
  {} & \multicolumn{4}{c}{\textbf{PES-WEKBP}} & \multicolumn{4}{c}{\textbf{Comparison method}} \\ 
\hline
 Input size &  {\makecell{20$\times $20:\\(27, 48)}}  & {\makecell{30$\times $30:\\(13, 24)}}
  & {\makecell{40$\times $40:\\(10, 19)}}
  & {\makecell{50$\times $50:\\(10, 19)}}
  & 256, 256  & 128, 128  & 64, 64  & 32, 32 \\
 Iterations number  &  50  & 50 & 50 & 70
  & 80  & 80  & 150  & 150 \\
  {\makecell{Minimum loss\\ during training}}  &  1.0149  & 1.1755 & 1.1529 & 1.1937 & 0.9722  & 0.9980  & 1.0393  & 1.1796 \\
  {\makecell{Top-1\\training accuracy}}  &  0.5596  & 0.5116 & 0.5315 & 0.5374 & 0.5775  & 0.5473  & 0.5453  & 0.4510 \\
  Training time (s)  &  1.12  & 1.11 & 1.07 & 1.04 & 1491  & 1198  & 935  & 880 \\
  {\makecell{$PEC{I_{{\rm{Tr}}}}$\\(Top-1, $\alpha  = 0.5$)}}  &  0.7092  & 0.6440 & 0.7012 & 0.7251 & 0.0713  & 0.0662 & 0.0709  & 0.0540 \\
   Test time (s)  & 1.23  & 1.20 & 1.31 & 1.19 & 6.88 & 5.89 & 5.00  & 4.51 \\
  {\makecell{Top-1\\prediction accuracy}}  & 0.6189  & 0.6180 & 0.6067 & 0.6368 & 0.5951 & 0.5985 & 0.5692  & 0.5549\\
  {\makecell{$PEC{I_{{\rm{Te}}}}$\\(Top-1, $\alpha  = 0.5$)}}  & 0.7433  & 0.7599 & 0.6816 & 0.7942 & 0.2800 & 0.2950 & 0.3065  & 0.3211\\
\toprule[1pt]
\end{tabular}
\end{center}
\end{table*}  

Table \ref{Accuracytab} compares the Top-$ G $ beam prediction accuracy and iteration counts between PES-WEKBP and the baseline under different input sizes. Combined with the accuracy curves in Fig. \ref{BeamPrePlot}, the results confirm that PES-WEKBP maintains higher prediction accuracy with smaller data volumes. 
PES-WEKBP demonstrates superior accuracy across evaluation metrics when using 50-unit block processing at $10 \times 19$ resolution. The method achieves 63.68\% Top-1 accuracy, an improvement of 14.76\% relative to the baseline's 55.49\% with $32 \times 32$ inputs. 
This performance advantage extends to Top-4 evaluation, where PES-WEKBP reaches 99.74\% accuracy.
Additionally, PES-WEKBP achieves convergence in just 70 iterations with 50-unit blocks compared to the baseline's 150 iterations using $32 \times 32$ inputs. This 53\% reduction in iteration count highlights the method's effectiveness in minimizing model complexity through material and location-aware WEK's integrated material reflection and distance feature processing.

The PCEI combines accuracy with a logarithmic transformation of training time and incorporates a time stability penalty, avoiding biases from linear relationships.
Table \ref{tab57} compares the Top-1 beam prediction accuracy, computational complexity, and PCEI metrics between PES-WEKBP and the baseline. 
Results show that PES-WEKBP achieves improvements exceeding three orders of magnitude in both training and inference times. Specifically, for $ B = 50 $, the average training time per iteration is reduced from 1491 seconds to 1.04 seconds, representing a speedup of approximately 1400 times.
Inference time for PES-WEKBP (1.19 s) is 5.8 times faster than the baseline (6.88 s). In terms of PCEI, PES-WEKBP scores 0.7942 in testing, 2.47 times higher than the baseline’s 0.3211, demonstrating superior real-time performance.
This efficiency stems from material and location-aware WEK’s redundancy compression while preserving critical EM propagation features through material-weight mapping.

\section{Conclusion}\label{FIVE}
This paper presents PES-WEKBP, a novel architecture that integrates visual sensing with wireless environment knowledge for high-precision, low-complexity channel prediction task. 
The framework uniquely combines radio propagation theory, computer vision, and neural networks to address critical challenges of data redundancy and training inefficiency in conventional approaches. Our key innovation, the material and location-aware WEK method, transforms visual information into EM-relevant semantics by preserving propagation-critical features while eliminating redundant elements, establishing quantitative relationships between material properties, spatial configurations, and wave propagation characteristics.
In beam prediction tasks, the proposed solution achieves remarkable efficiency gains, reducing input dimensions by 99.75\% to 99.96\% and improving accuracy by 5.52\% to 8.19\% compared to environment semantic-based methods. We further contribute the PCEI metric, which innovatively evaluates model performance through accuracy-time trade-off analysis using logarithmic transformation and exponential decay terms, effectively addressing linearity bias and training instability. Experimental validation confirms the superior PCEI performance of our approach across diverse operational scenarios.

\section{Acknowledgment}
This work is supported by the National Science Fund for Distinguished Young Scholars (No.61925102), and the National Natural Science Foundation of China (No.62525101, 62401084, 62201087).

\bibliographystyle{IEEEtran}
\bibliography{bibsample}


\vspace{12pt}

\end{document}